\documentclass{optica-article}

\usepackage{graphicx}  
\graphicspath{{data/v1_images_red/}, {data/v2_images/},  {data/v3_image/}, {data/v3_images-3/}, {data/v3_images-4/}, {data/v3_images-5/}, {images/}, {data/ivcurve/}, {data/current_modeling/}, {data/compare_squares/}}
\DeclareGraphicsExtensions{.pdf,.png,.jpg}
\usepackage{wrapfig} 

\journal{opticajournal} 

\articletype{Research Article}

\usepackage{lineno}

\begin{document}

\title{Investigation of the Effects of the Multiplication Area Shape on the Operational Parameters of InGaAs/InAlAs SPADs}

\author{Anton Losev,\authormark{1,2,3} Alexandr Filyaev,\authormark{1,2,4,*}, Vladimir Zavodilenko,\authormark{1,2,4}, Igor Pavlov,\authormark{1,2} and Alexander Gorbatsevich\authormark{3,5}}

\address{\authormark{1}"QRate" LLC, St. Novaya 100, Moscow region, Odintsovo, Skolkovo, 143026, Russia\\
\authormark{2}National University of Science and Technology MISIS, Leninsky prospect 4, Moscow, 119333, Russia\\
\authormark{3}National Research University of Electronic Technology MIET, Shokin Square 1, Zelenograd, 124498, Russia\\
\authormark{4}HSE University, Myasnitskaya ulitsa 20, Moscow, 101000, Russia\\
\authormark{5}P.N. Lebedev Physical Institute of the Russian Academy of Sciences,  Leninsky prospect, 53, Moscow, 119333, Russia}

\email{\authormark{*}alex.filyaev.98@gmail.com} 


\begin{abstract*} 
A 2D model of an InGaAs/InAlAs single photon avalanche photodiode has been developed. The influence of the active area structure in the multiplication region on the diode's operating parameters has been studied. It was found that changing the diameter of the structure's active region leads to a change in the dark current in the linear part of the current-voltage curve and a change in the breakdown voltage. Reducing the diameter of the active region from 25 $\mu$m to 10 $\mu$m allowed decreasing the dark current in the linear mode by about $10$ dB. It has been shown that the quality of the SPAD device can be assessed by knowing the avalanche breakdown voltage and the overall current-voltage curve plot if we consider structures with the same multiplication region thickness and different remaining layers. The higher the breakdown voltage, the better the structure's quality due to smaller local increases in the field strength. Following this statement, we conclude that for further use in single-photon detectors, it is reasonable to pick specific SPADs from a batch on the sole basis of their current-voltage curves.
\end{abstract*}

\section{Introduction}
A single-photon avalanche photodiode (SPAD) for detecting emission at a wavelength of  $\lambda = 1550$ nm can be fabricated from different materials. For example, structures based on InGaAs/InP semiconductor materials have found widespread application \cite{signorelli2021, Signorelli2022, Liang2022, Kizilkan2022, Robinson2022, Baek2021, Sanzaro2018, WANG20191134, Bimbova2022, Jiang2018}. Diodes based on this material pair demonstrate superior values of essential operational characteristics: high photon detection efficiency ($PDE$), low dark count rate ($DCR$), small afterpulse probability ($AP$), and relatively low dead time ($DT$).
  
However, the scientific community continues to develop SPADs based on other material pairs, such as Si/Ge  \cite{Kirdoda2019, Wanitzek2020, Chen2021, Dumas2019, DOLLFUS2022108361} and InGaAs/InAlAs \cite{Lee2021, Chang2019, Wu2021, Zhang2021, Zhang2022}. In this paper, we study an InGaAs/InAlAs structure.

The feature of SPAD based on InGaAs/InAlAs is that the avalanche generation process is excited by a single electron, unlike the InGaAs/InP structure, where it is triggered by a hole. Accordingly, the avalanche process is mainly supported by the represented types of carriers in the avalanche \cite{Cao2019, Zheng2018}.

The advantage of using InAlAs material as a multiplication region is the high electron mobility, which allows much quicker quenching of the excited avalanche and bringing the structure to its equilibrium state. As a result, this device can have a higher cut-off frequency and better afterpulse characteristics than a device with an InP multiplication region. This is because the hole mobility in InP is lower than the electron mobility in InAlAs \cite{Chen2018}. 

Nevertheless, commercially available SPAD devices for $\lambda = 1550$ are mainly built on InGaAs/InP materials. The main obstacle to the fabrication of SPADs based on In- GaAs/InAlAs materials is a large number of defects in the InAlAs material, which does not allow benefiting from realizing its full potential \cite{Jiang2019, Sim2021}. Improvements in the sputtering processes of this material will enable competing with these two types of devices. 

In addition, the optimal sequence of heterojunction layers and their parameters are likely to differ significantly between InGaAs/InP and InGaAs/InAlAs-based devices. Structure optimization, which is addressed in this work, is the second vital task to unlock the potential of the InGaAs/InAlAs device.

In this work, the effect of the shape of the multiplication region on the electric field strength distribution and the avalanche generation rate in the structure has been investigated. The current-voltage characteristic (CVC) plots for different types of structures were also compared.

\section{2D modeling of InGaAs/InAlAs SPADs structure}

The modeling was performed in the T-CAD system. In the simulation, specific carrier lifetime values were set as parameters to calculate the recombination rate by the Schockley-Reed-Hall (SRH) mechanism. Radiative and Auger recombination processes were also included in the calculation. The finite element modeling solved a system of continuity equations for electrons and holes and Poisson's equation.

In order to limit the active region of the multiplication region (the region where the main avalanche generation process takes place), it is necessary to use different widths of the multiplication region in the active and inactive multiplication regions. In the active region, we need to achieve a higher field strength, so the width of the forbidden region should be smaller. The following problems arise when implementing this principle of building an active region.

\begin{itemize}
    \item One has to arrange a transition between the two widths of the multiplication region. This can be done using either a sharp or a smooth transition. However, any sharp transition will result in a local increase in intensity and, therefore, a significant increase in the dark count rate when the instrument is operating in Geiger mode.
    \item More than one transition can be used to minimize the field strength in the inactive region. This means that not only two different multiplication region widths can be used, but three or more. This solution enables a significant reduction of the volume of high electric field areas in the inactive region and, consequently, a substantial decrease in the dark count rate when the device is operated in Geiger mode.
\end{itemize}

In this paper, three SPAD structures with different shapes of the multiplication region have been proposed. The first structure had two levels of the multiplication region and a sharp transition (2 lvls sharp), the second one had three levels of the multiplication region and a sharp transition (3 lvls sharp), and the third one had three levels of the multiplication region and a smooth transition (3 lvls smooth). The width of the multiplication region in the active region of all structures was the same: 0.8 $\mu$m. The width of the buffer 2 region in the active region was also the same: 1.8 $\mu$m. The thickness of each level of the multiplication region was 0.6 $\mu$m. 

In the simulated structure, the diameter of the active region is  25 $\mu$m. The diameter of the whole simulated structure was 45 $\mu$m. 

In the following subsections, the profiles of the electric field strength distribution and the avalanche generation rate in each of the described structures are considered.

\subsection{Two-level multiplication region with a sharp transition}

\begin{figure}[ht!]
    \centering
    \includegraphics[width=7cm]{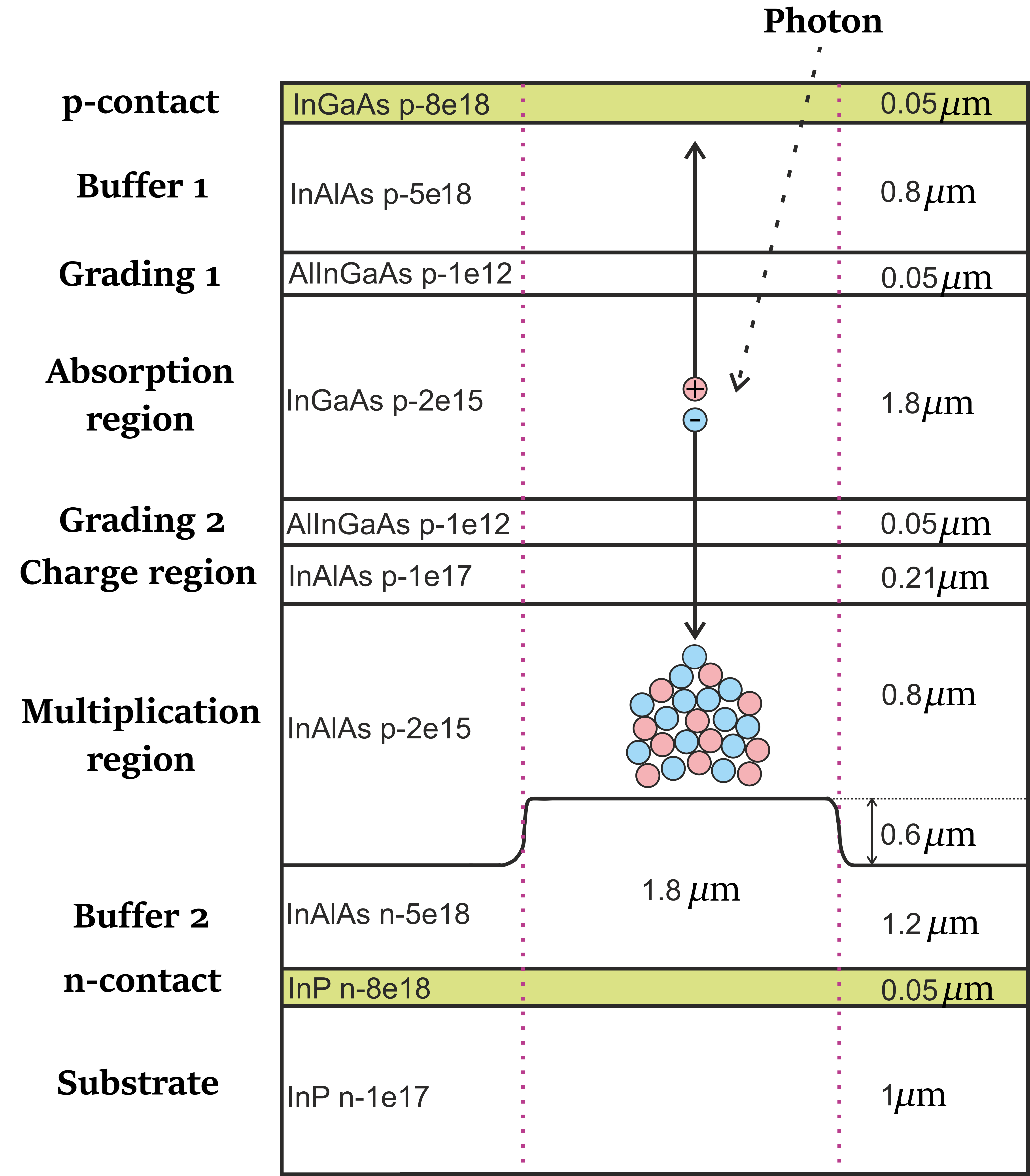}
    \caption{Schematic representation of a 2D SPAD structure with a two-level multiplication region with a sharp transition.}
    \label{fig:2d_struct_2sharp}
\end{figure}

A structure with two levels of multiplication region with a sharp transition is shown in Figure \ref{fig:2d_struct_2sharp}. The dotted line separates the active and inactive regions.

\begin{figure}[ht!]
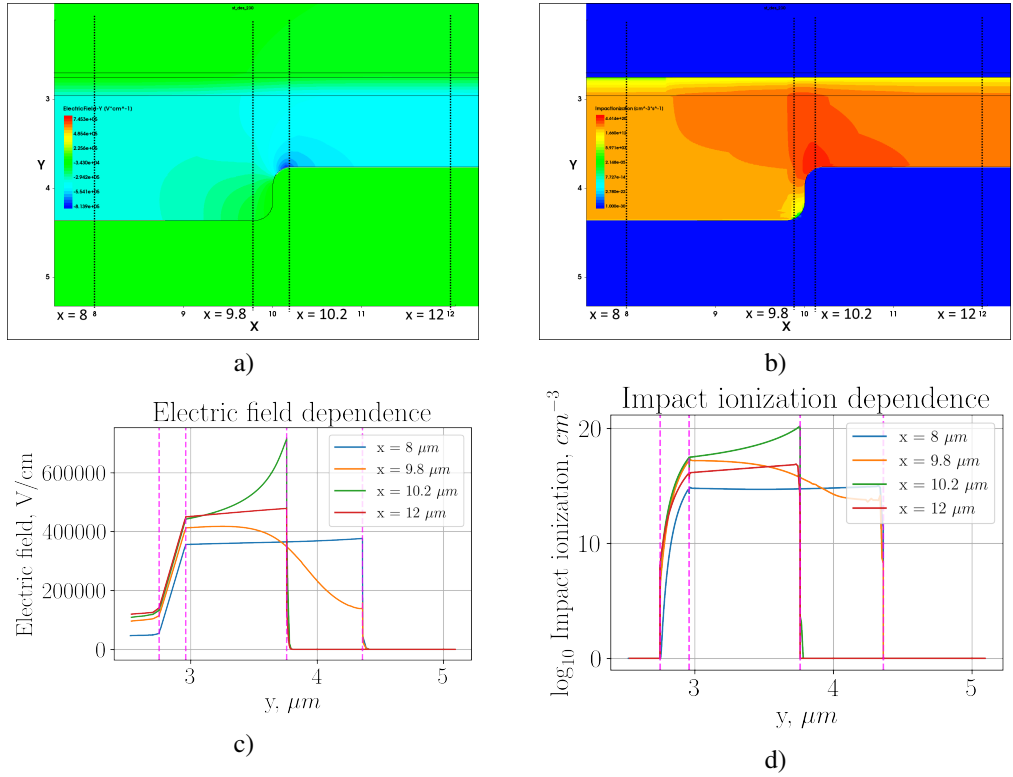

    \begin{minipage}[h]{0.47\linewidth}
    \center{\includegraphics[width=1\linewidth]{v1_elfy_230}} a) \\
    \end{minipage}
    \hfill
    \begin{minipage}[h]{0.47\linewidth}
    \center{\includegraphics[width=1\linewidth]{v1_impion_230}} \\b)
    \end{minipage}
    \vfill
    \begin{minipage}[h]{0.47\linewidth}
    \center{\includegraphics[width=1\linewidth]{v1_elf_230_py}} c) \\
    \end{minipage}
    \hfill
    \begin{minipage}[h]{0.47\linewidth}
    \center{\includegraphics[width=1\linewidth]{v1_impion_230_py}} d) \\
    \end{minipage}
    \caption{Two-level multiplication region with sharp transition: a) heat map of the electric field strength distribution; b) heat map of the avalanche generation rate distribution; c) electric field distribution profile in the cross-sections indicated in figure a); d) avalanche generation rate distribution profile in the cross-sections indicated in figure b).}
    \label{fig:2ur_sharp}
\end{figure}

The Figure \ref{fig:2ur_sharp} shows heat maps of electric field distribution and avalanche generation rate (graphs a) and b), respectively) in the multiplication region, as well as profiles of these parameter distributions in individual sections (graphs c) and d), respectively).

The disadvantage of using a sharp transition is a large local increase in the electric field strength, as shown in graph c). For the section $x = 10.2 \ \mu m$, the electric field strength increased to values of $E_{loc-enh} = 700 \ kV / cm$, which is $40 \%$ stronger than the field in the remaining active region $E_{act} = 500 \ kV / cm$. This field increase led to an increase in the avalanche generation rate in this local area of about 60 dB. Since the size of this region is about 0.5 $\mu$m, we can estimate the contribution of this detriment to the overall $DCR$. The ring area where this effect occurs can be calculated as  $S_{loc-enh} = {\pi}/{4} \ (D^2_{loc-enh+} - D^2_{loc-enh-}) = 0.785 * (25.5^2 - 24.5^2) \approx 40 \ \ \ \mu m^2$. It takes about  $10 \ \%$ of the active area. Accordingly, the fraction of the dark counts generated by the local increase in field strength will be 40 dB greater than in the active region. It is, therefore, more appropriate to focus development efforts directly on addressing this drawback.

In the next section, where three multiplication regions with sharp transitions are considered, an attempt is made to reduce the  $DCR$ by reducing the field strength in the inactive region and hence the avalanche generation rate.

\subsection{Three-level multiplication region with a sharp transition}

\begin{figure}[ht!]
    \centering
    \includegraphics[width=0.5\linewidth]{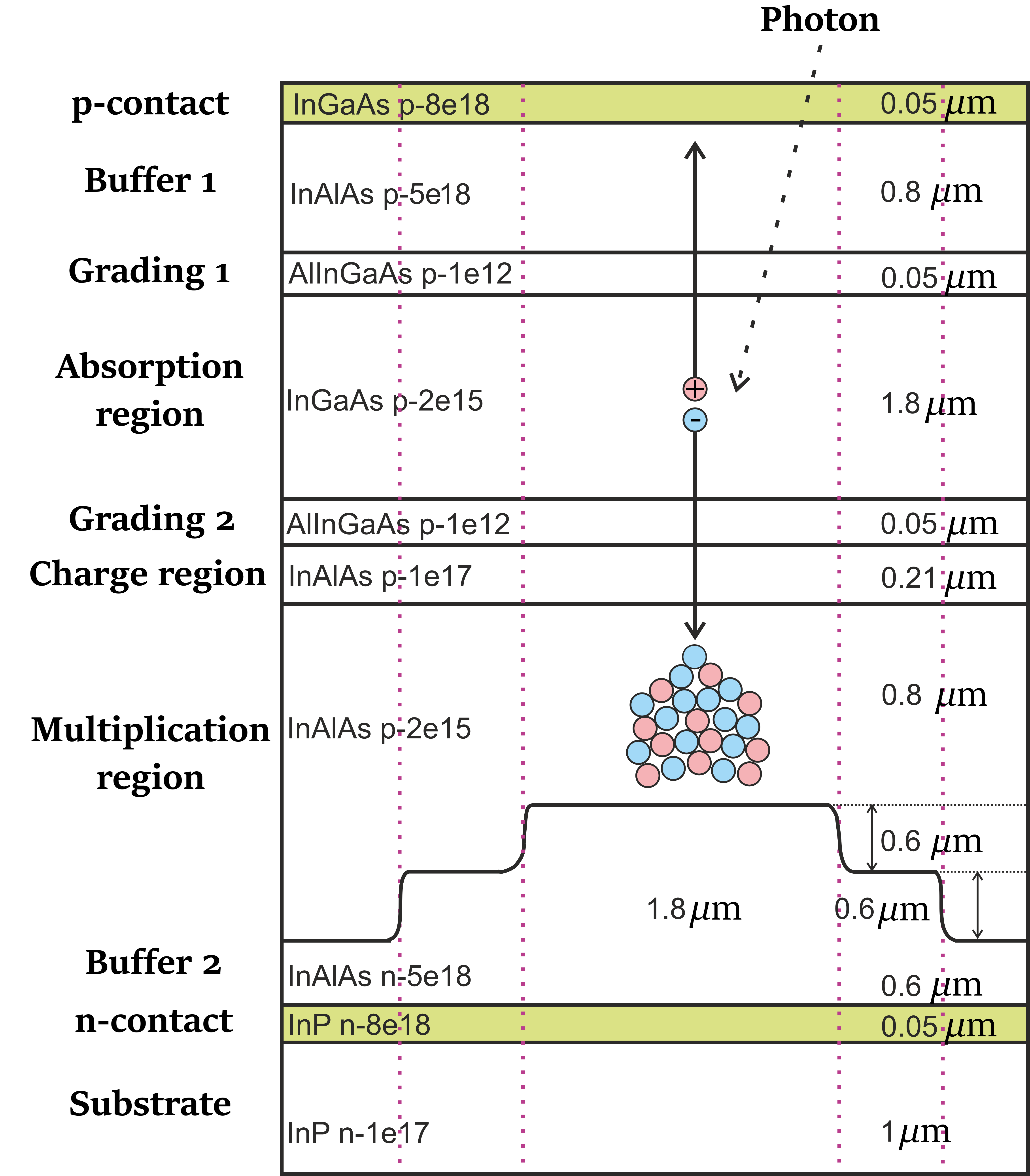}
    \caption{Schematic representation of a 2D SPAD structure with a three-level multiplication region with a sharp transition.}
    \label{fig:2d_struct_3sharp}
\end{figure}

A structure with three levels of multiplication region with a sharp transition is shown in Figure \ref{fig:2d_struct_3sharp}. The dotted line separates the active and inactive regions.

\begin{figure}[ht!]
    \begin{minipage}[h]{0.47\linewidth}
    \center{\includegraphics[width=1\linewidth]{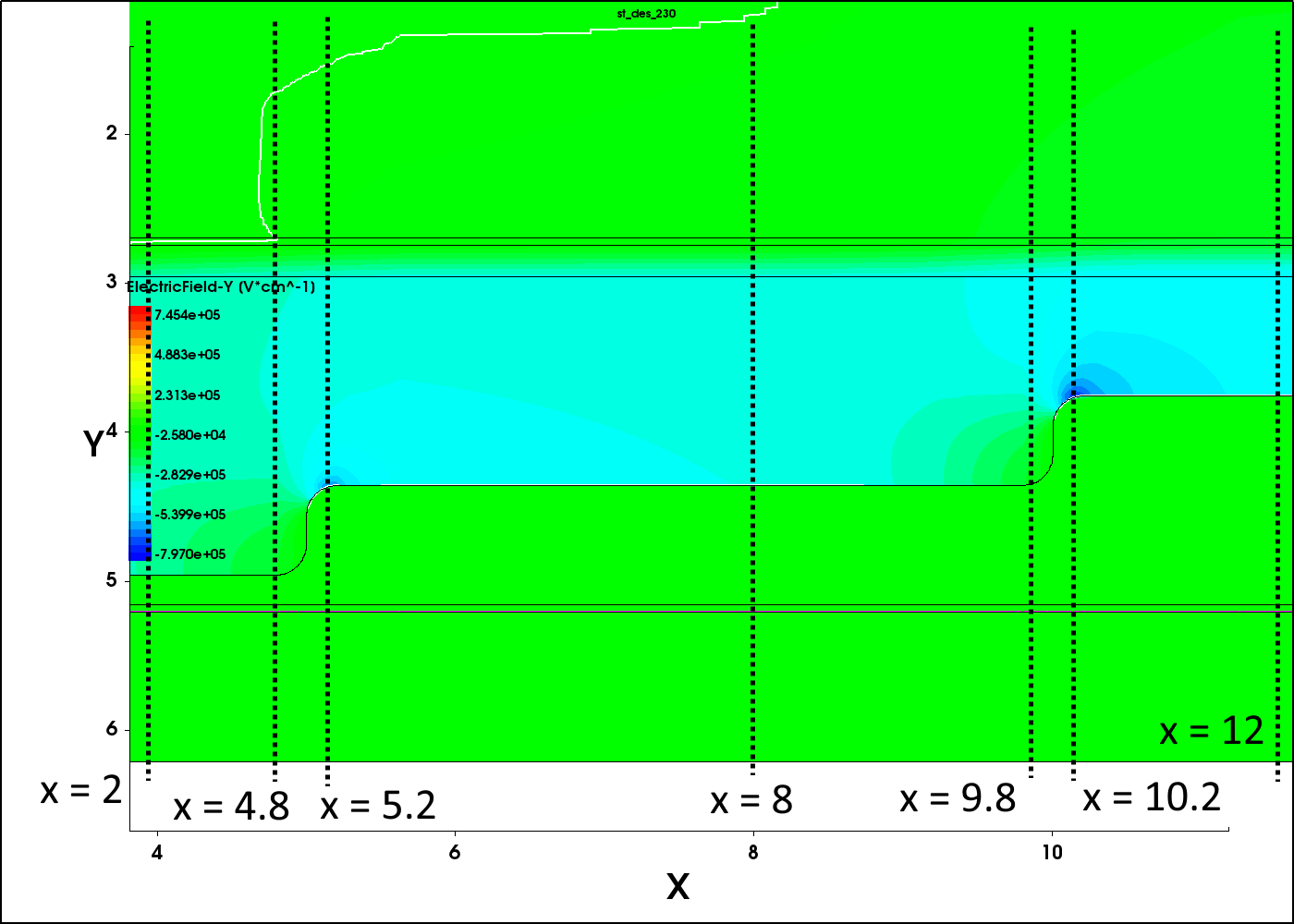}} a) \\
    \end{minipage}
    \hfill
    \begin{minipage}[h]{0.47\linewidth}
    \center{\includegraphics[width=1\linewidth]{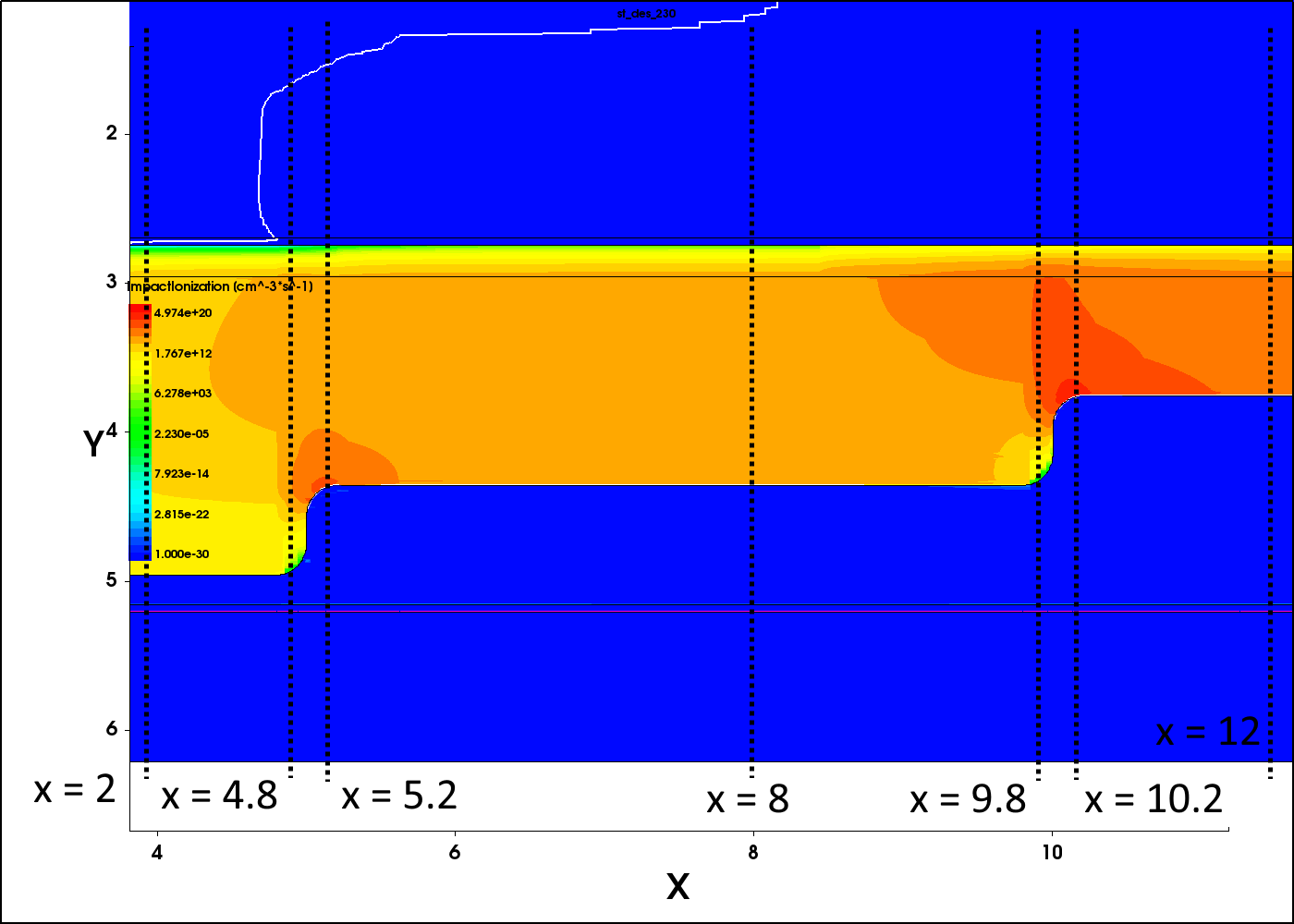}} \\b)
    \end{minipage}
    \vfill
    \begin{minipage}[h]{0.47\linewidth}
    \center{\includegraphics[width=1\linewidth]{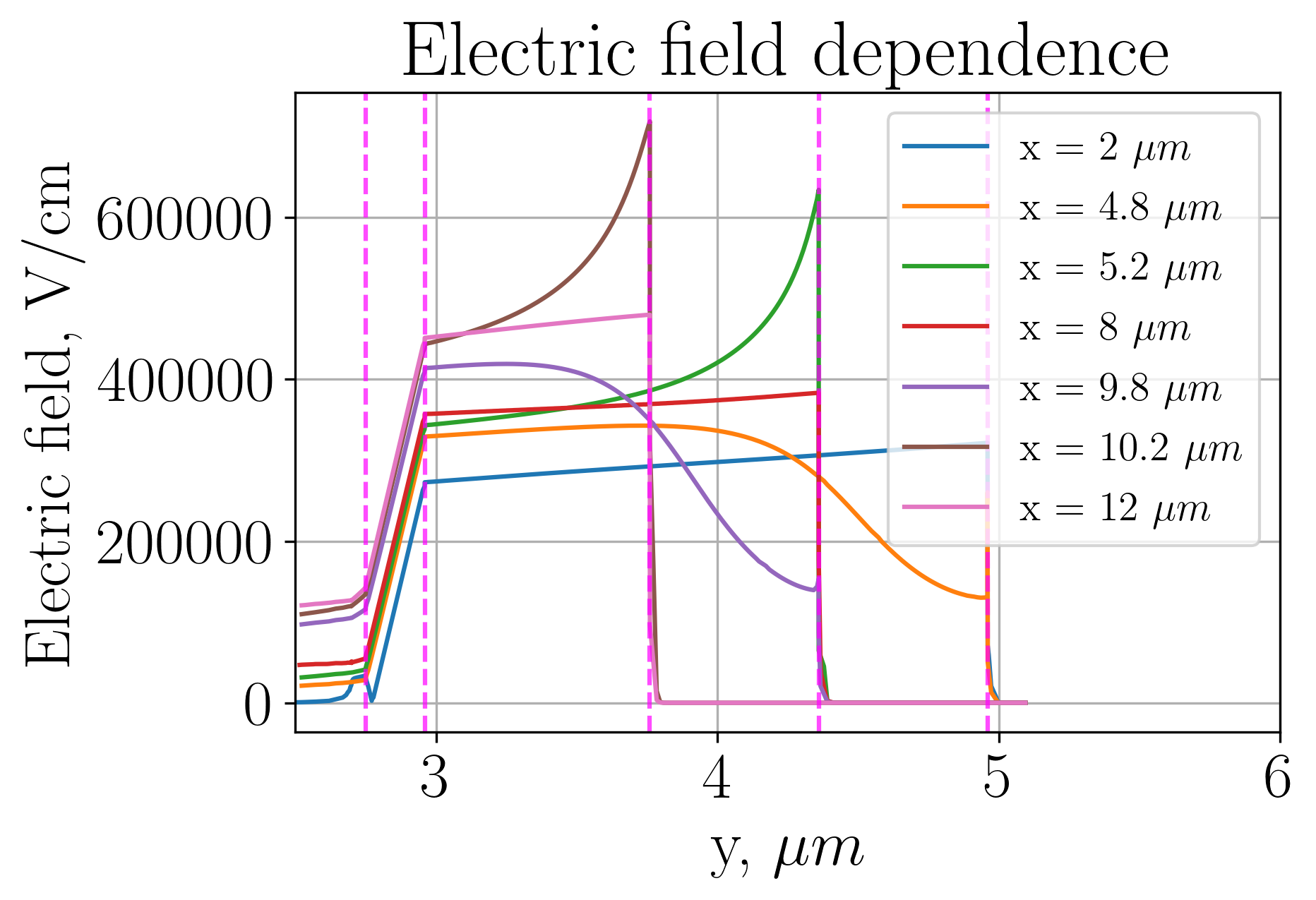}} c) \\
    \end{minipage}
    \hfill
    \begin{minipage}[h]{0.47\linewidth}
    \center{\includegraphics[width=1\linewidth]{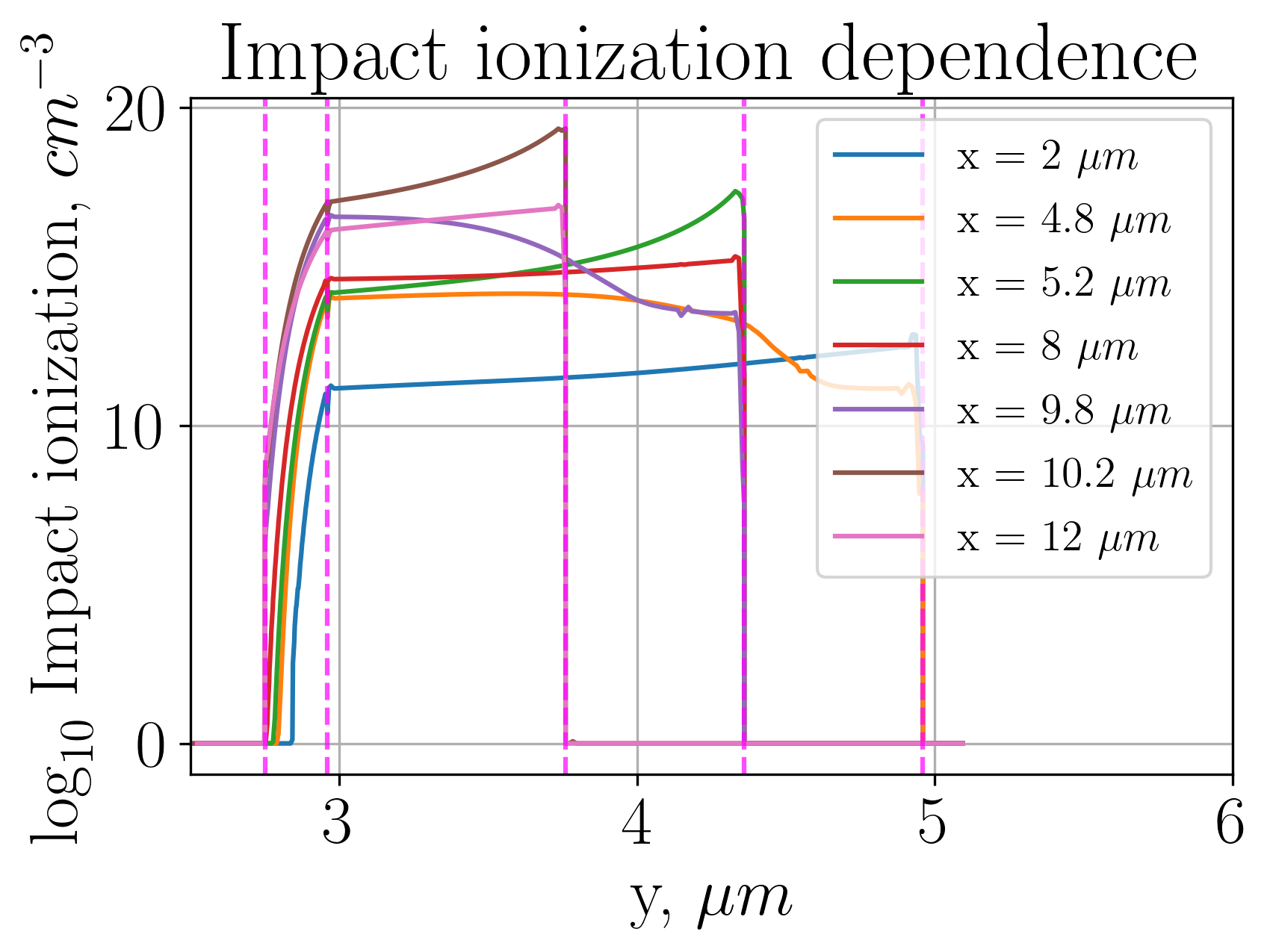}} d) \\
    \end{minipage}
    \caption{Three-level multiplication region with sharp transition: a) heat map of the electric field strength distribution; b) heat map of the avalanche generation rate distribution; c) electric field distribution profile in the cross-sections indicated in figure a); d) avalanche generation rate distribution profile in the cross-sections indicated in figure b).}
    \label{fig:3ur_sharp}
\end{figure}

The Figure \ref{fig:3ur_sharp} shows thermal maps of electric field distribution and avalanche generation rate (graphs a) and b), respectively, in the multiplication region, as well as profiles of these parameter distributions in individual sections (graphs c) and d), respectively).

The described optimization of the structure, on the contrary, has further aggravated the problem of the local increase of the electric field and the avalanche generation rate because another ring has been added (the outer one) in which we have again seen the effect of the local increase of the parameters at the sharp transition. As can be seen in Figure \ref{fig:3ur_sharp} c) and d), the electric field strength in the second ring reaches $620 \ kV / cm$ and the avalanche generation rate is about $20$ dB higher than in the active region. This value does not make such a large additional contribution to the total dark count rate compared to the first transition. However, as can be seen, in the thickest multiplication region ($x = 2 \ \ \mu m$) the avalanche generation rate is about four orders of magnitude lower than in the active region. In the mid-thickness multiplication region, the avalanche generation rate is about two orders of magnitude lower than in the active region.

Thus, the use of a three-level multiplication region structure is only justified if the problem of increased local field at sharp transitions can be solved.

\subsection{Three-level multiplication region with a smooth transition}

\begin{figure}[ht!]
    \centering
    \includegraphics[width=0.5\linewidth]{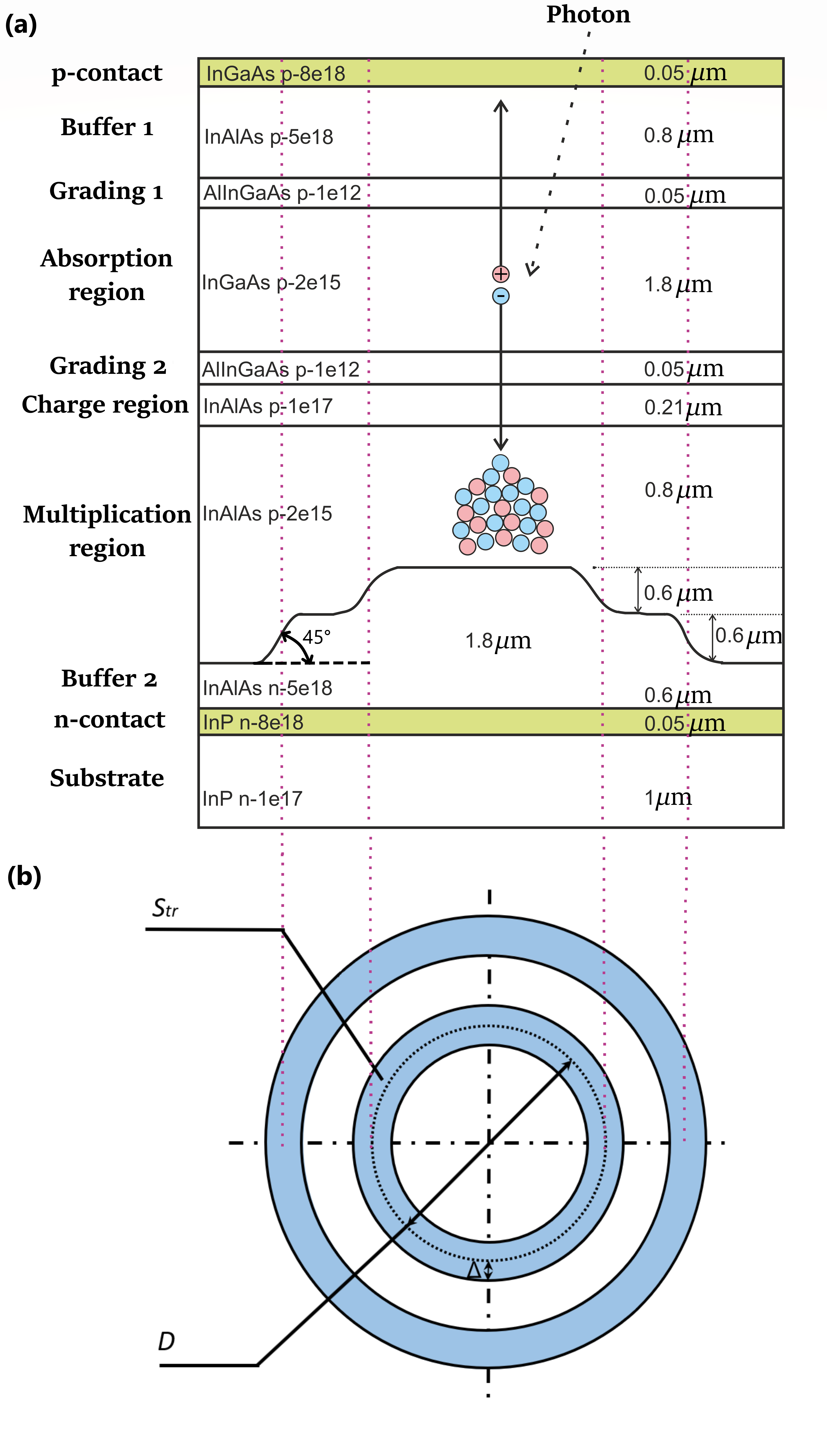}
    \caption{ a) Schematic representation of a 2D SPAD structure with a three-level multiplication region with smooth transition; b) schematic representation of smooth transitions in the multiplication area: $\Delta$ - half-width of the transition area; $D$ - diameter of the active area, $S_{tr}$ - area of the transition area.}
    \label{fig:2d_struct_3smooth}
\end{figure}

A structure with three levels of multiplication region with a smooth transition is shown in Figure \ref{fig:2d_struct_3smooth}. The dotted line separates the active and inactive regions.

\begin{figure}[ht!]
    \begin{minipage}[h]{0.47\linewidth}
    \center{\includegraphics[width=1\linewidth]{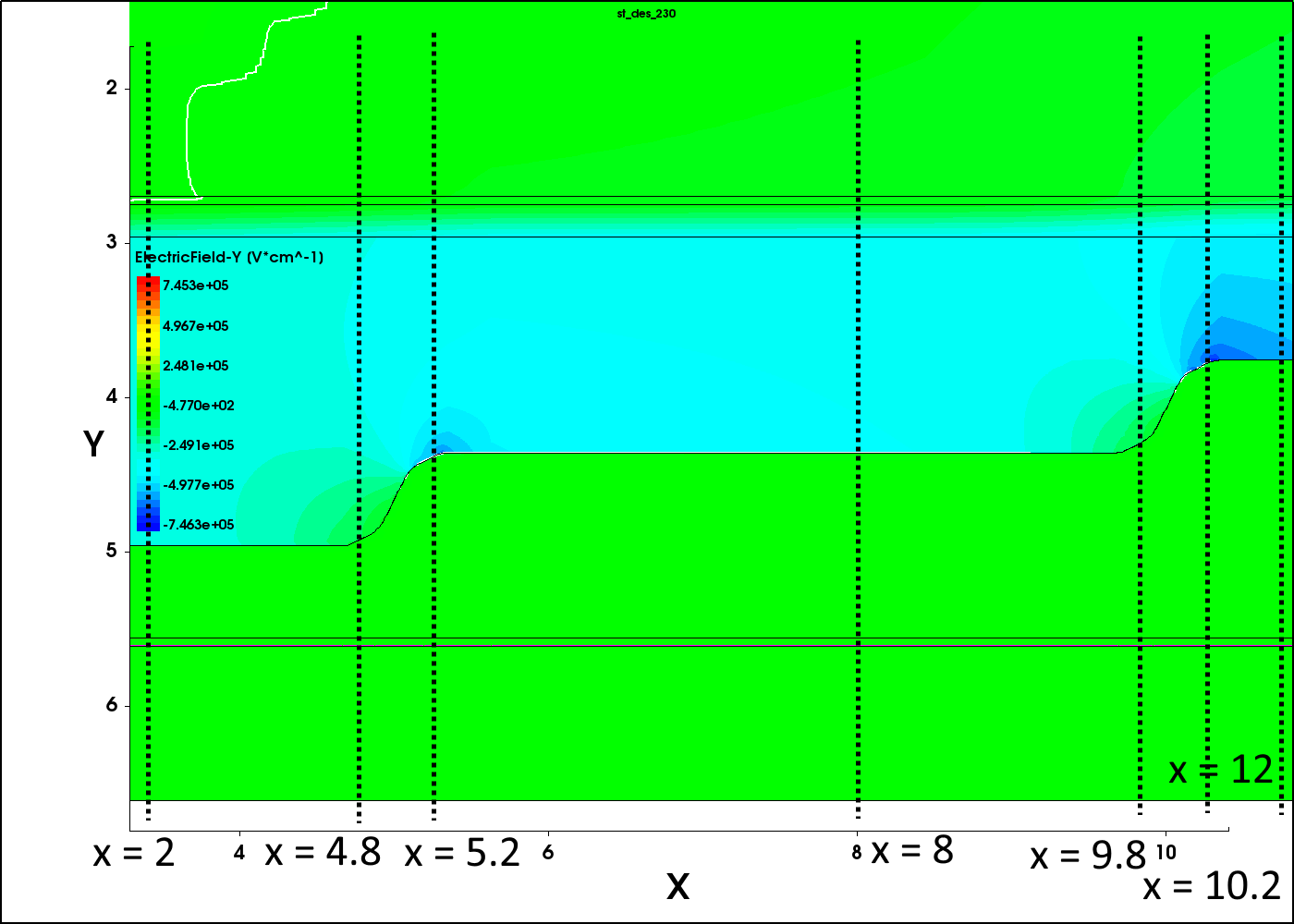}} a) \\
    \end{minipage}
    \hfill
    \begin{minipage}[h]{0.47\linewidth}
    \center{\includegraphics[width=1\linewidth]{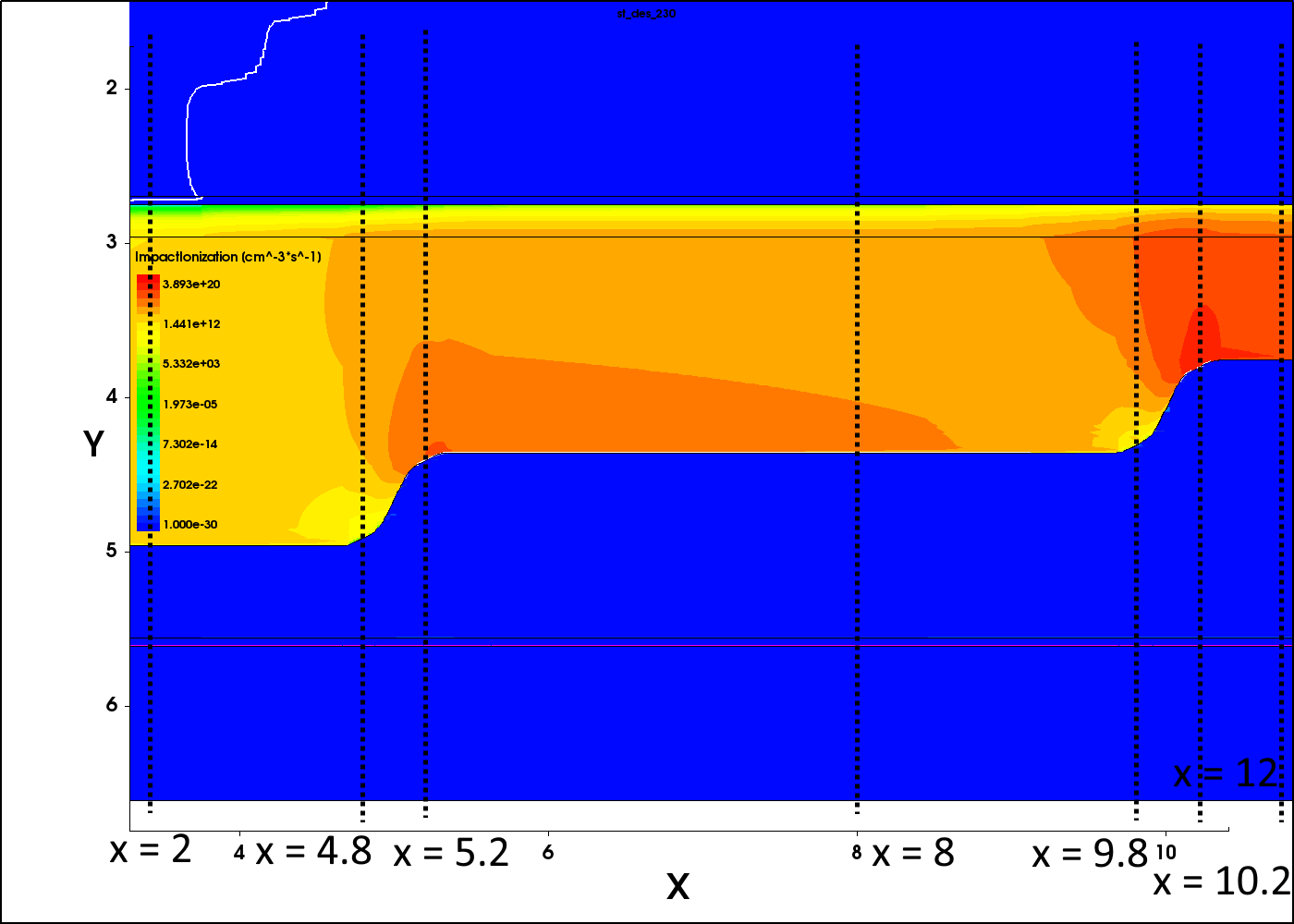}} \\b)
    \end{minipage}
    \vfill
    \begin{minipage}[h]{0.47\linewidth}
    \center{\includegraphics[width=1\linewidth]{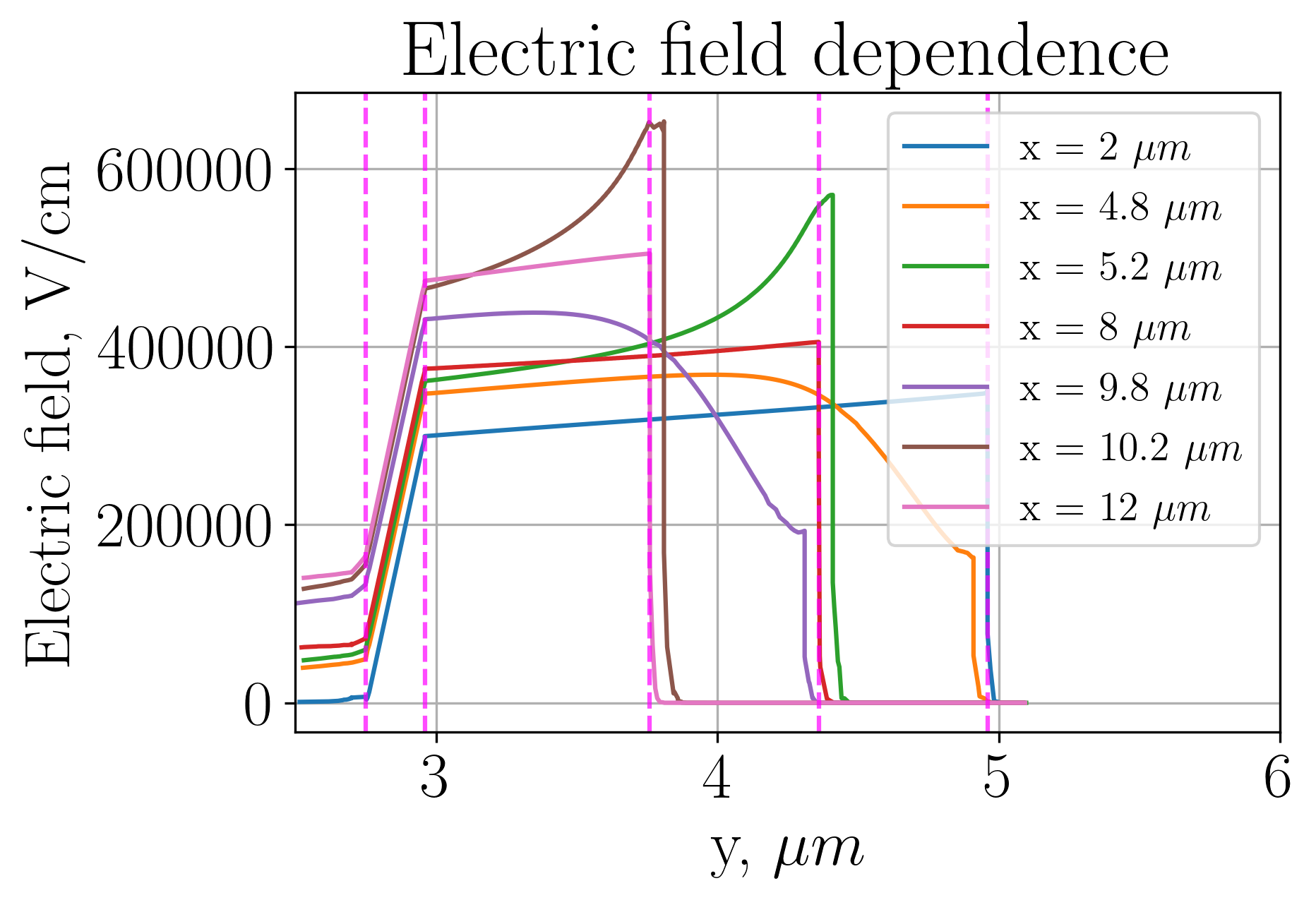}} c) \\
    \end{minipage}
    \hfill
    \begin{minipage}[h]{0.47\linewidth}
    \center{\includegraphics[width=1\linewidth]{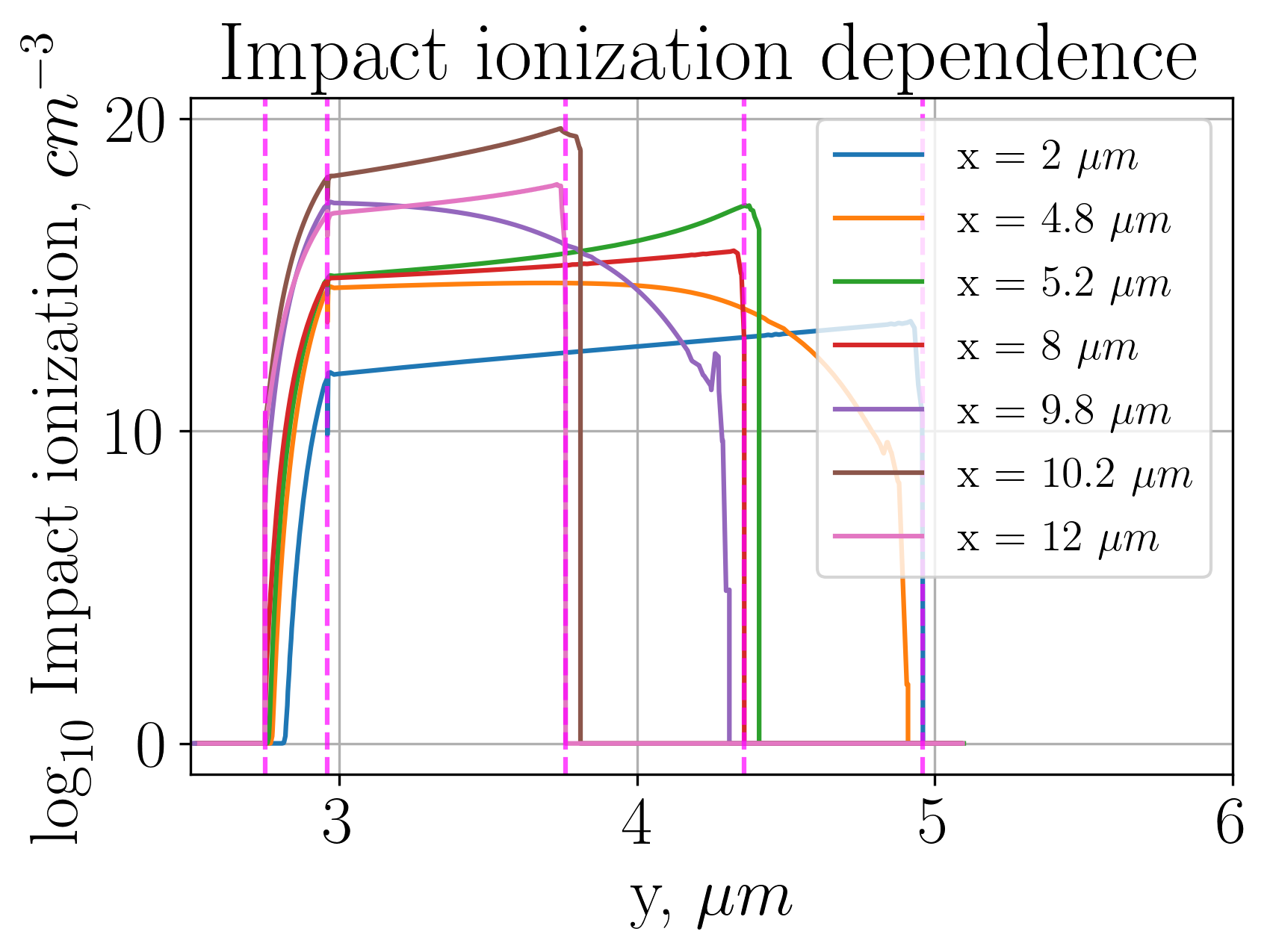}} d) \\
    \end{minipage}
    \caption{Three-level multiplication region with smooth transition: a) heat map of the electric field strength distribution; b) heat map of the avalanche generation rate distribution; c) electric field distribution profile in the cross-sections indicated in figure a); d) avalanche generation rate distribution profile in the cross-sections indicated in figure b). }
    \label{fig:3ur_smooth}
\end{figure}

Figure \ref{fig:3ur_smooth} shows heat maps of electric field distribution and avalanche generation rate (graphs a) and b), respectively) in the multiplication region, as well as profiles of these parameter distributions in individual sections (graphs c) and d) respectively).

Using a smooth transition has reduced the electric field strength in the region of maximum local enhancement to $620 \ kV/cm$ from $700 \ kV/cm$ with a sharp transition. It has reduced the ratio of the avalanche generation rate in the local enhancement region to that in the active region by two orders of magnitude. Thus, the dark count rate due to local enhancement is now only  $20$ dB $DCR$ in the entire active region. The avalanche generation rate in the second transition is approximately equal to the avalanche generation rate in the active region, and, accordingly, the dark count contribution is approximately $10 \%$ of the dark count rate in the active region. The dark count rate was reduced by nearly the same amount by using a three-level multiplication region, as compared to the two-level region structure.

Thus, creating smooth transitions at the boundaries of the multiplication region levels is an extremely effective method of reducing the dark count rate in the device as a whole. If one could achieve sufficiently smooth transitions (see Figure \ref{fig:2d_struct_3smooth}) in the design of the device so that the dark count rate contribution by the local enhancements is no more than $20$ dB higher than in the active region of the device, then a three-level multiplication region can be considered for implementation.  

\begin{figure}[ht!]
    \begin{minipage}[h]{0.47\linewidth}
    \center{\includegraphics[width=1\linewidth]{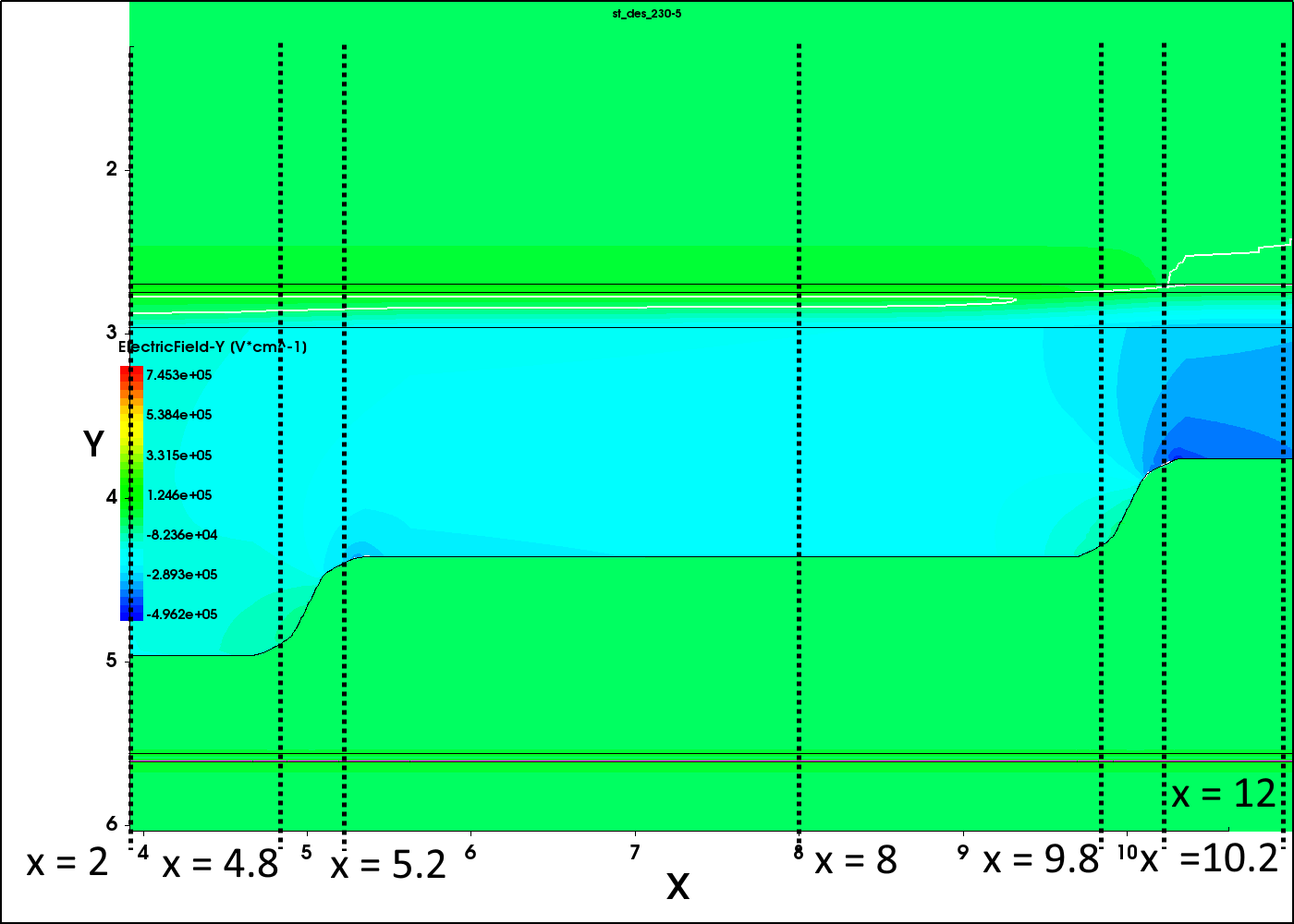}} a) \\
    \end{minipage}
    \hfill
    \begin{minipage}[h]{0.47\linewidth}
    \center{\includegraphics[width=1\linewidth]{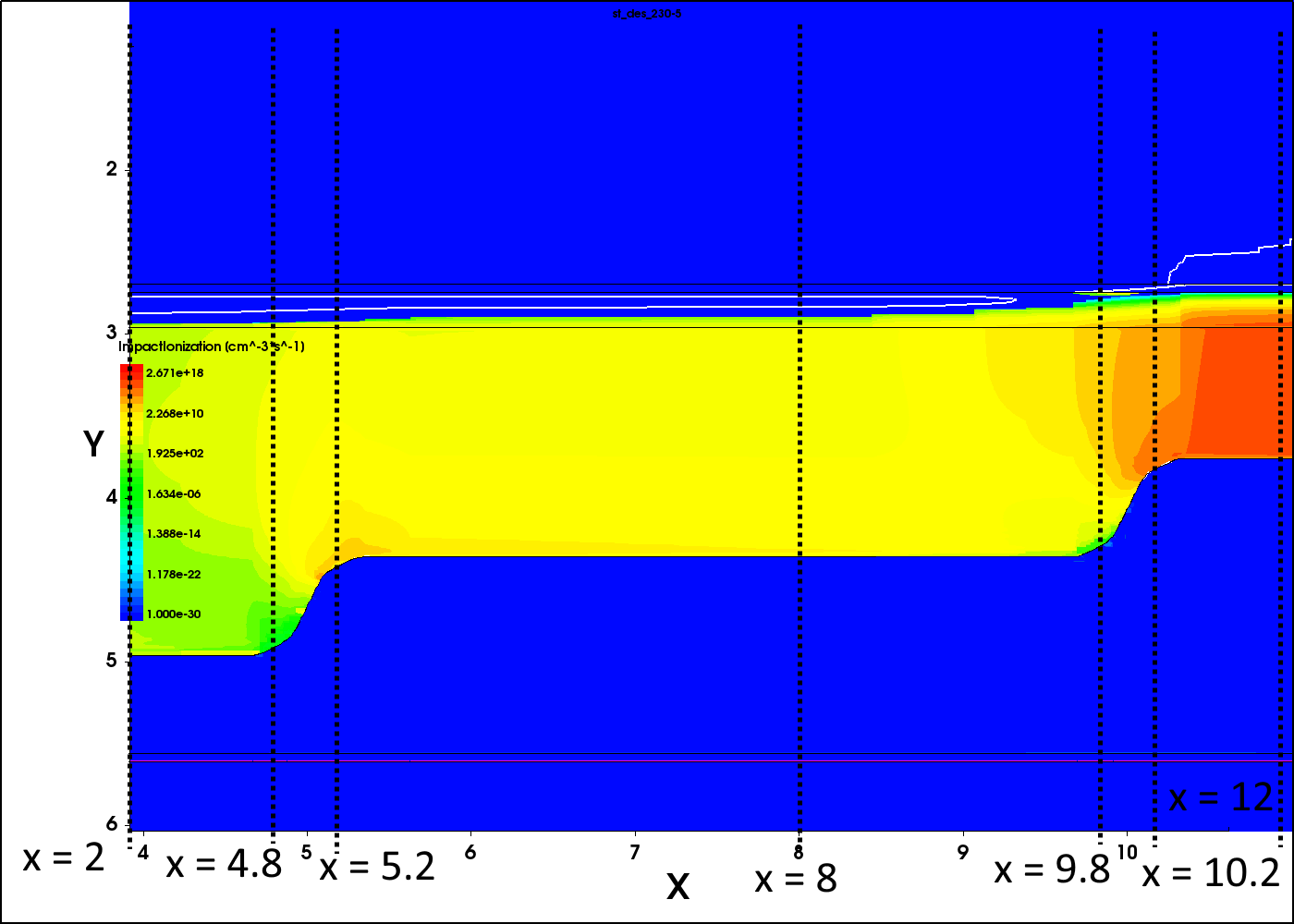}} \\b)
    \end{minipage}
    \vfill
    \begin{minipage}[h]{0.47\linewidth}
    \center{\includegraphics[width=1\linewidth]{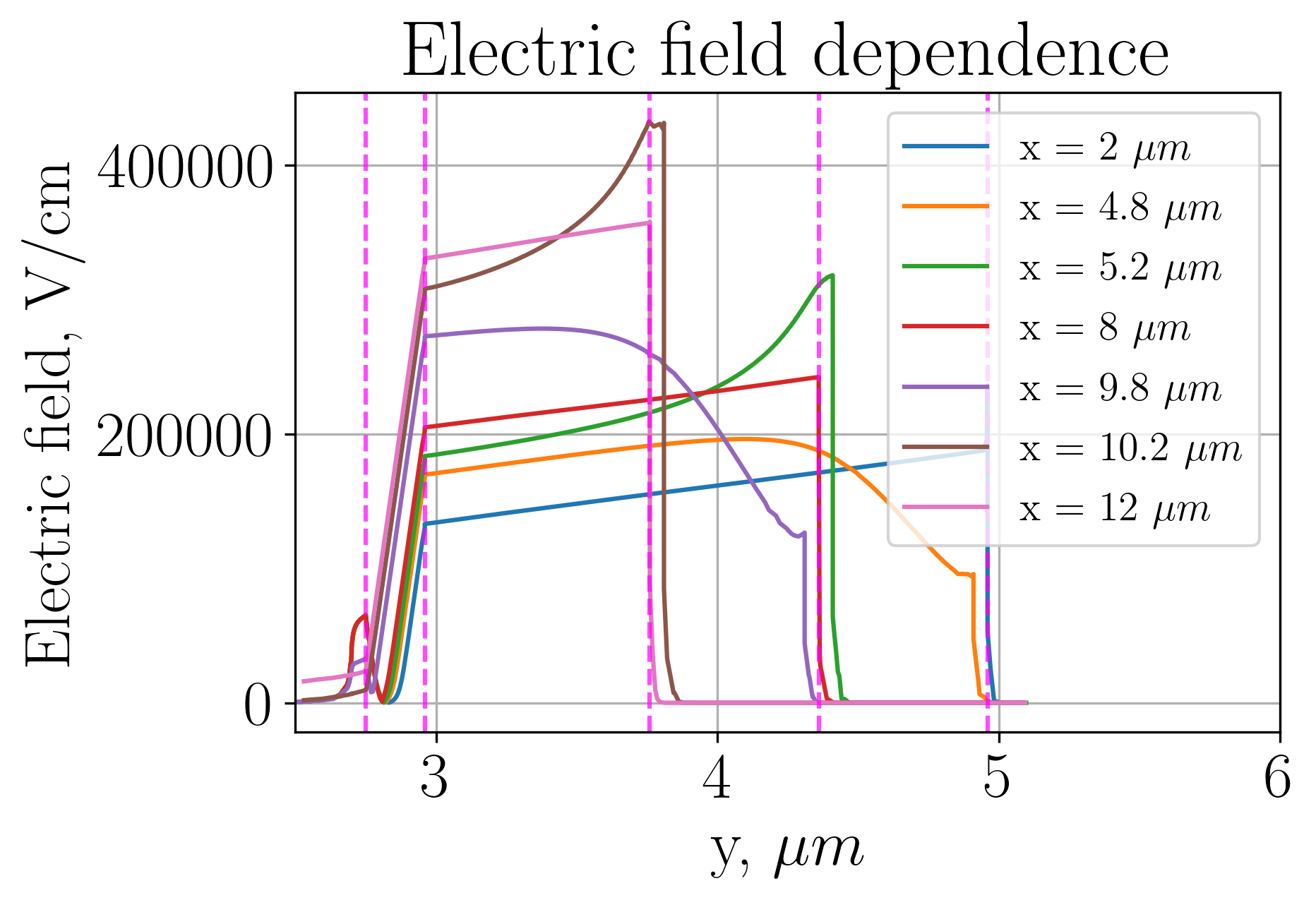}} c) \\
    \end{minipage}
    \hfill
    \begin{minipage}[h]{0.47\linewidth}
    \center{\includegraphics[width=1\linewidth]{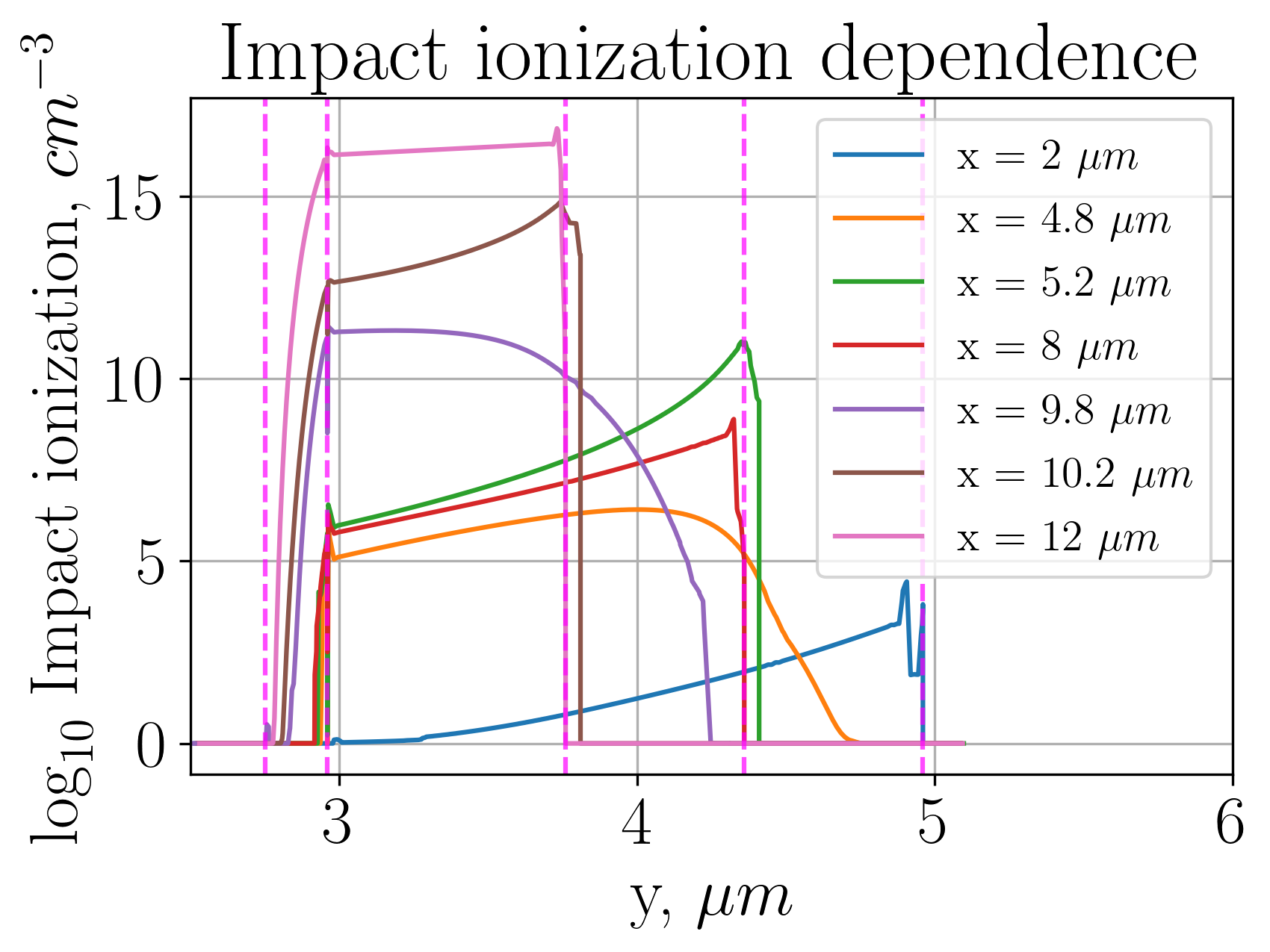}} d) \\
    \end{minipage}
    \caption{Three-level multiplication region structure with smooth transition with incident radiation intensity $10 \ \mu W / cm^2$: a) heat map of electric field strength distribution; b) heat map of avalanche generation rate distribution; c) electric field distribution profile in the cross-sections indicated in figure a); d) avalanche generation rate distribution profile in the cross-sections indicated in figure b).}
    \label{fig:3ur_smooth-5}
\end{figure}

\begin{figure}[ht!]
    \begin{minipage}[h]{0.47\linewidth}
    \center{\includegraphics[width=1\linewidth]{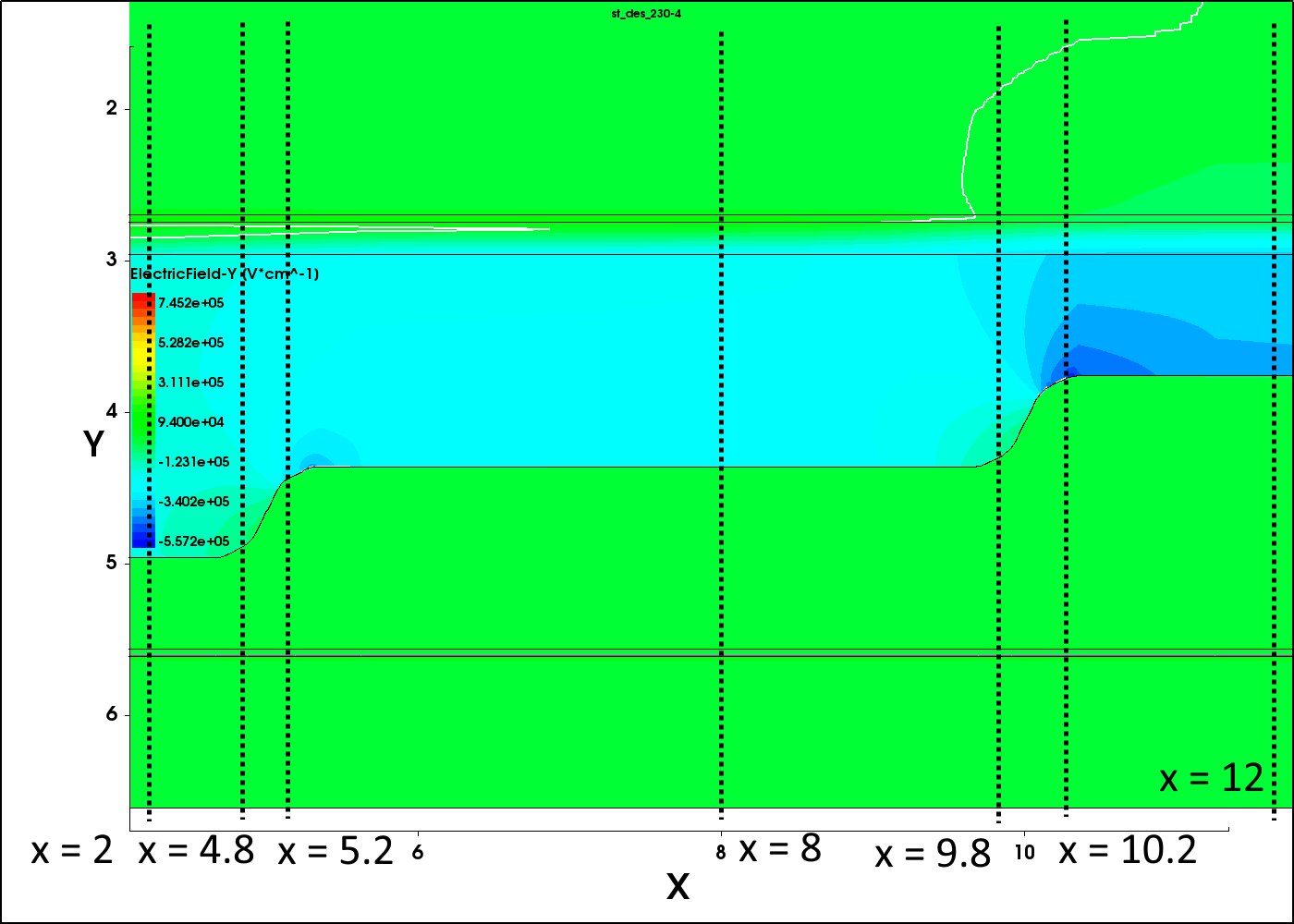}} a) \\
    \end{minipage}
    \hfill
    \begin{minipage}[h]{0.47\linewidth}
    \center{\includegraphics[width=1\linewidth]{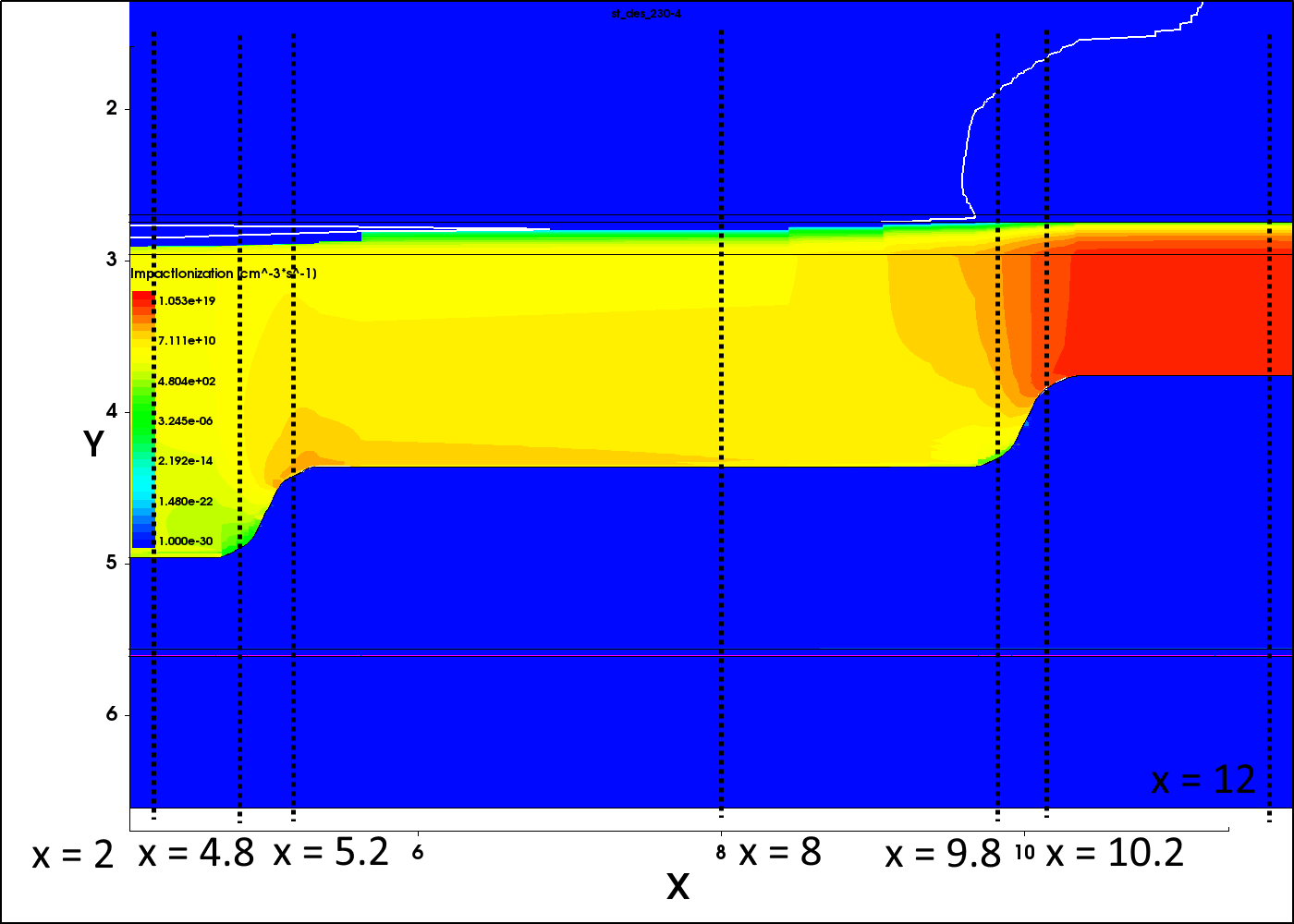}} \\b)
    \end{minipage}
    \vfill
    \begin{minipage}[h]{0.47\linewidth}
    \center{\includegraphics[width=1\linewidth]{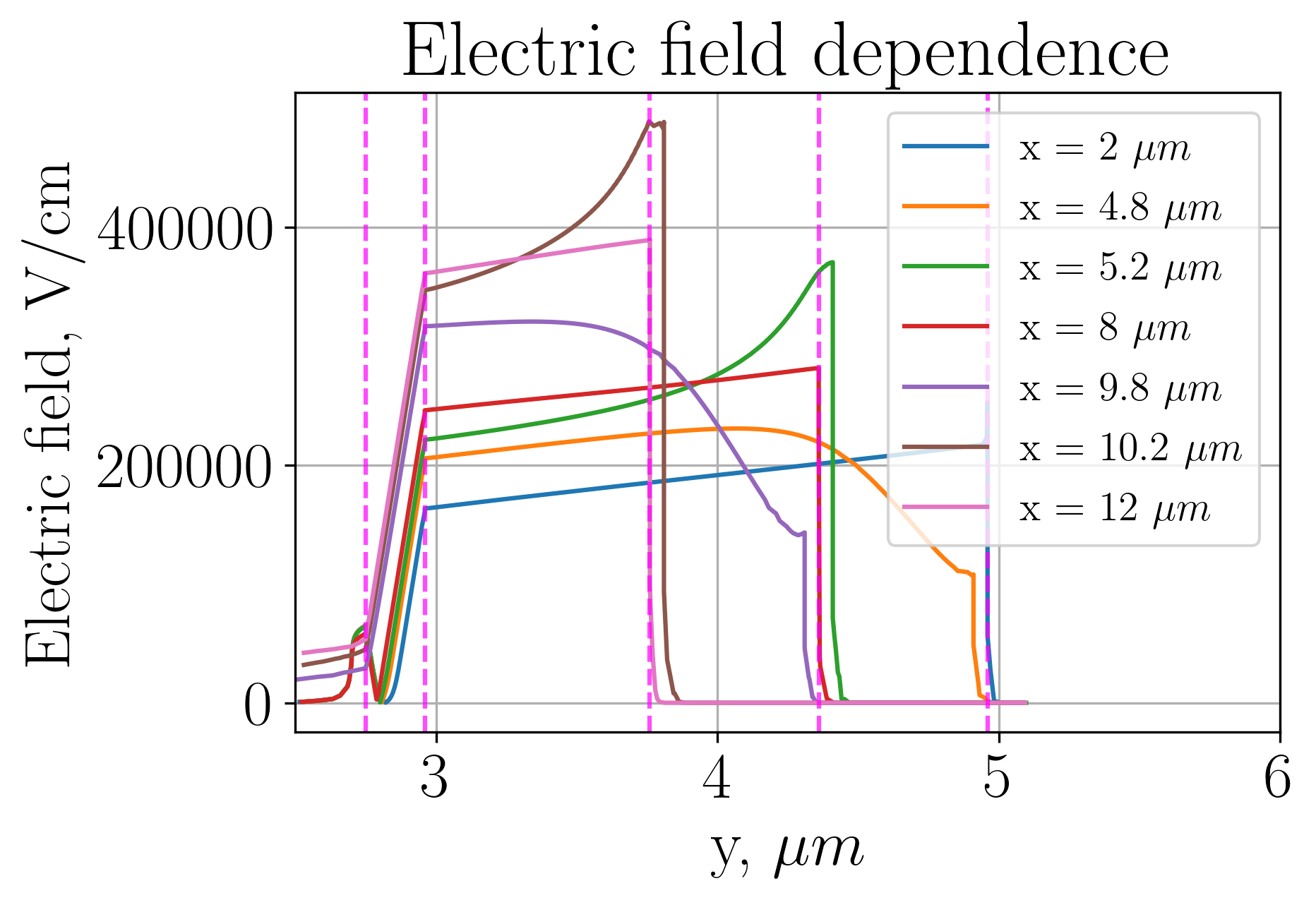}} c) \\
    \end{minipage}
    \hfill
    \begin{minipage}[h]{0.47\linewidth}
    \center{\includegraphics[width=1\linewidth]{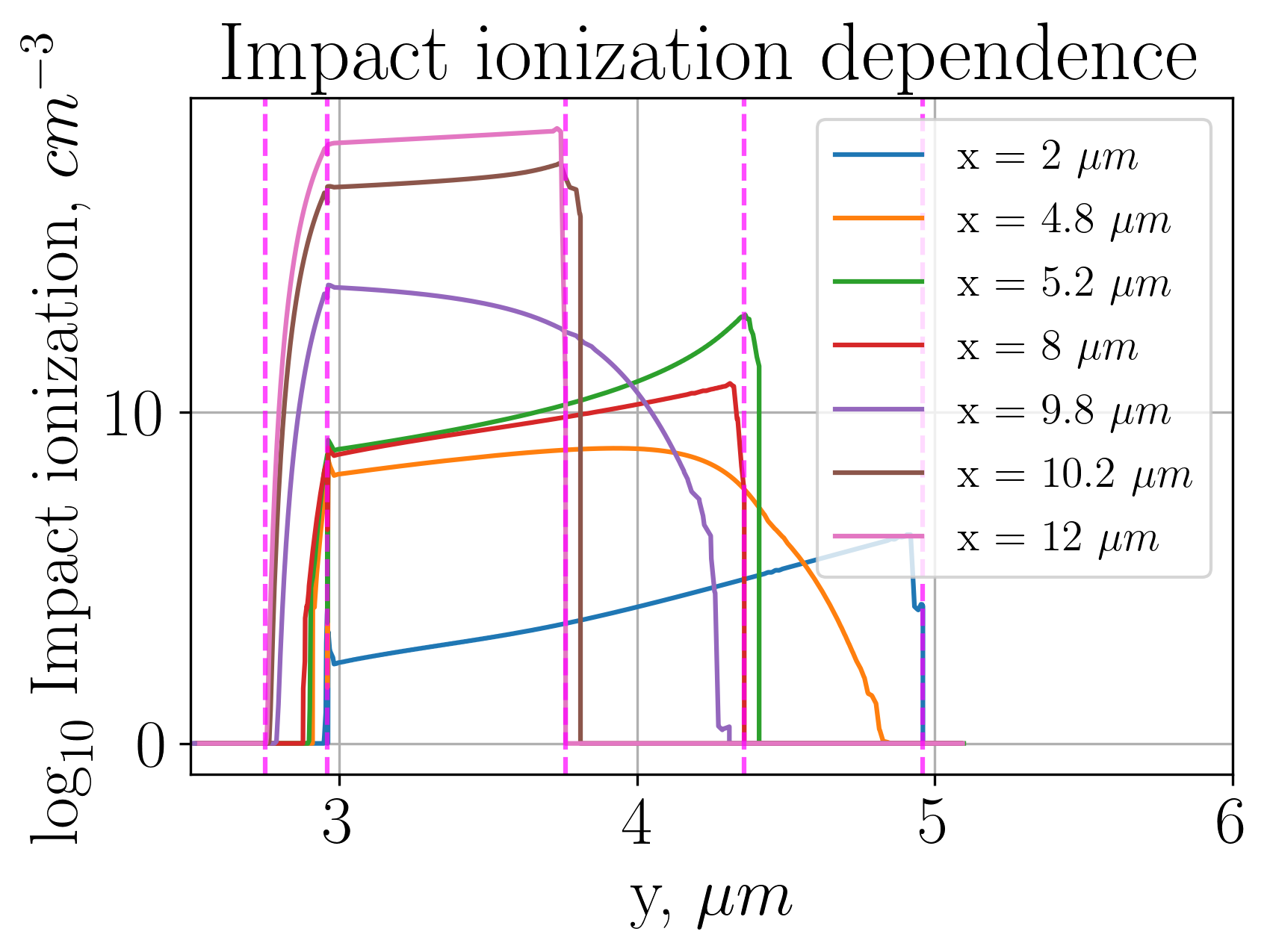}} d) \\
    \end{minipage}
    \caption{Three-level multiplication region structure with smooth transition with incident radiation intensity $100 \ \mu W / cm^2$: a) heat map of electric field strength distribution; b) thermal map of avalanche generation rate distribution; c) electric field distribution profile in the cross-sections indicated in figure a); d) avalanche generation rate distribution profile in the cross-sections indicated in figure b).}
    \label{fig:3ur_smooth-4}
\end{figure}

\begin{figure}[ht!]
    \begin{minipage}[h]{0.47\linewidth}
    \center{\includegraphics[width=1\linewidth]{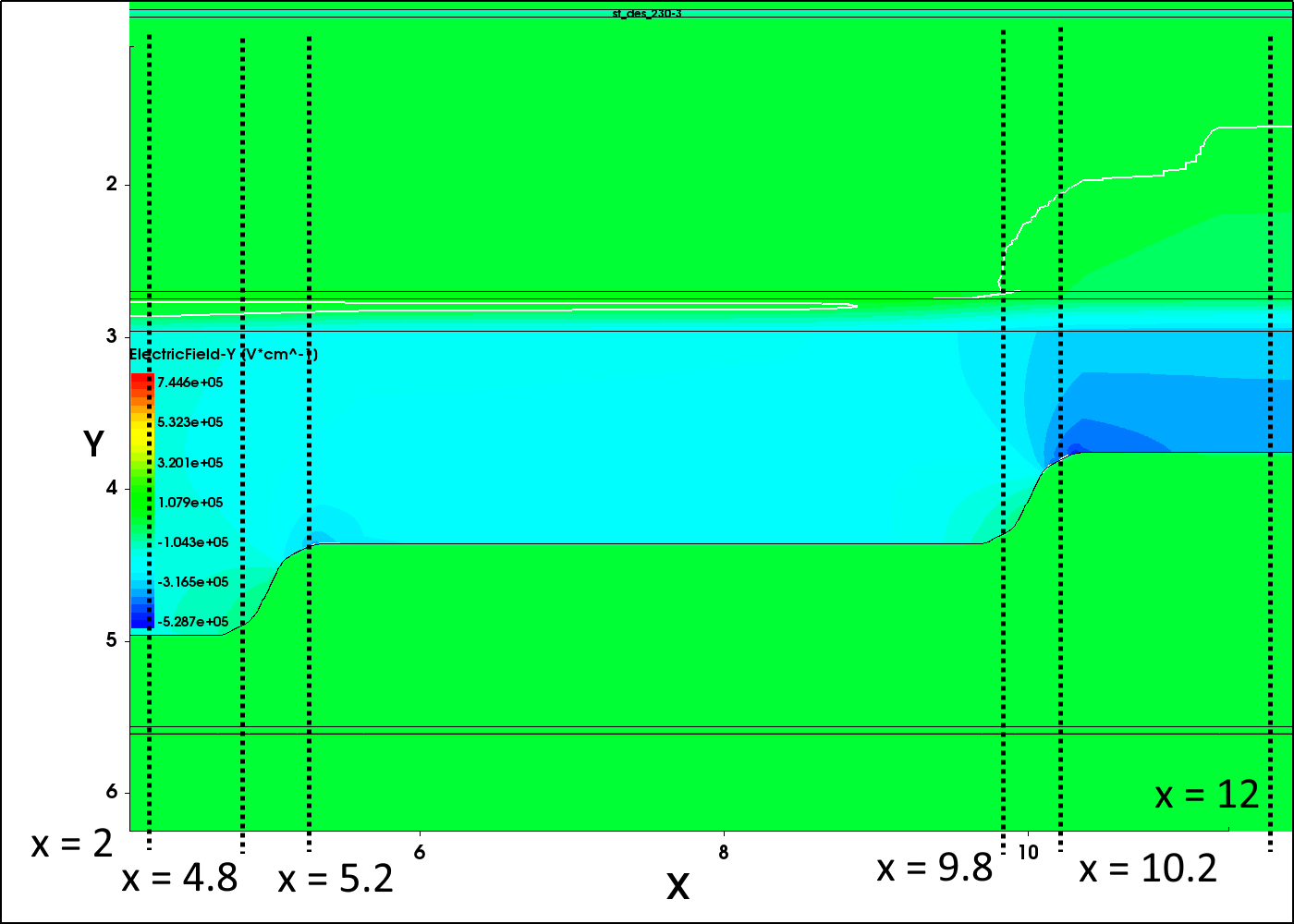}} a) \\
    \end{minipage}
    \hfill
    \begin{minipage}[h]{0.47\linewidth}
    \center{\includegraphics[width=1\linewidth]{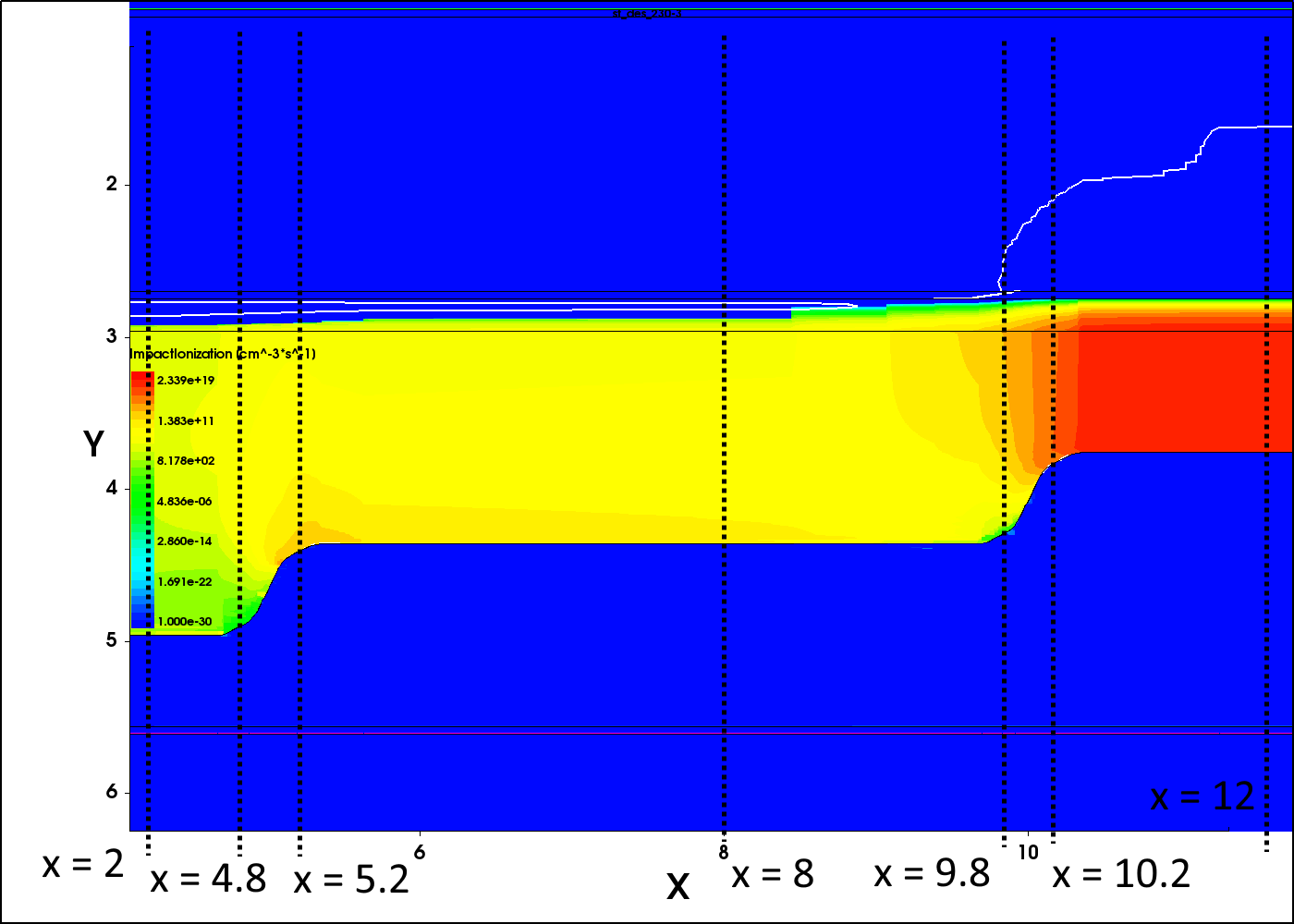}} \\b)
    \end{minipage}
    \vfill
    \begin{minipage}[h]{0.47\linewidth}
    \center{\includegraphics[width=1\linewidth]{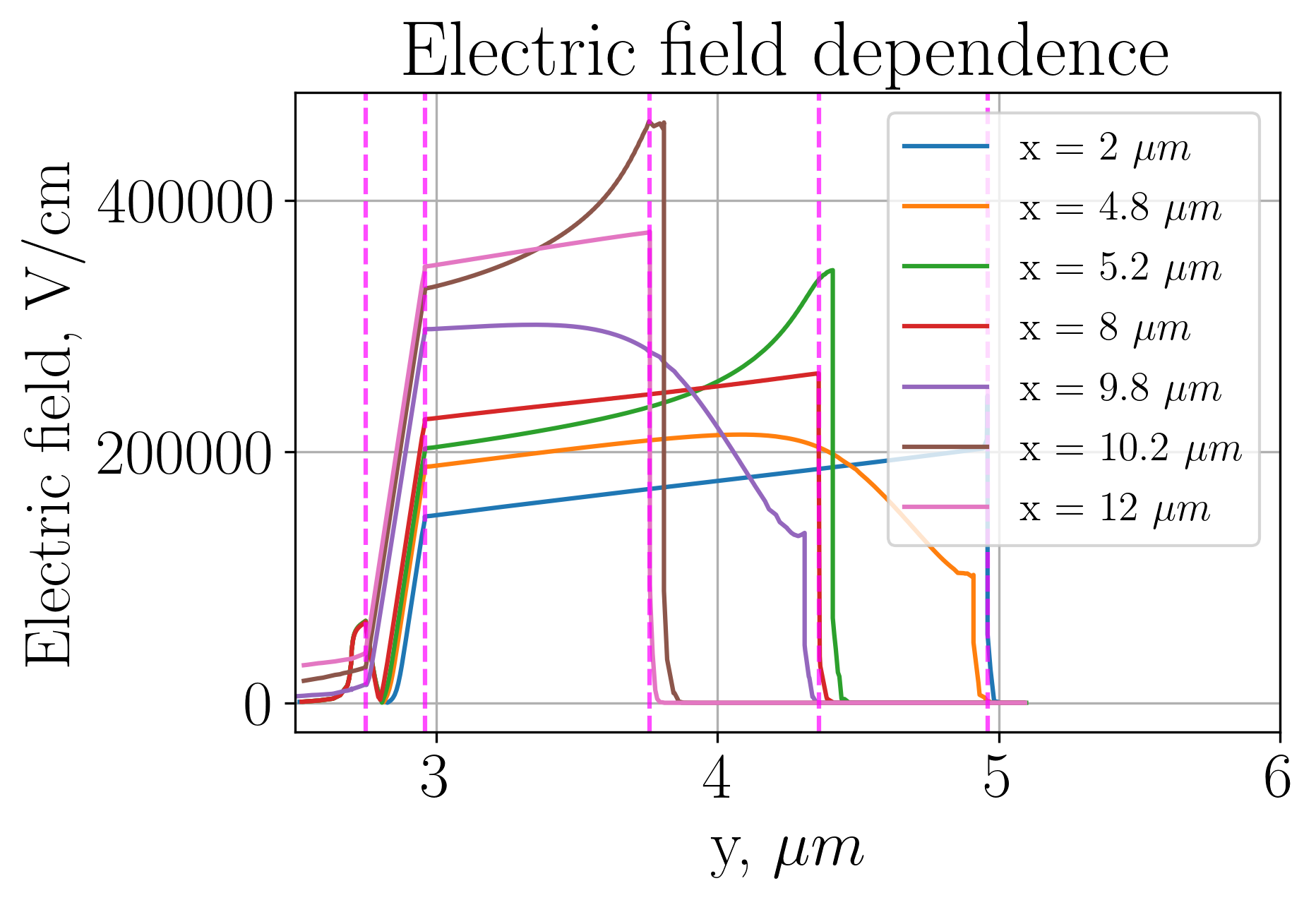}} c) \\
    \end{minipage}
    \hfill
    \begin{minipage}[h]{0.47\linewidth}
    \center{\includegraphics[width=1\linewidth]{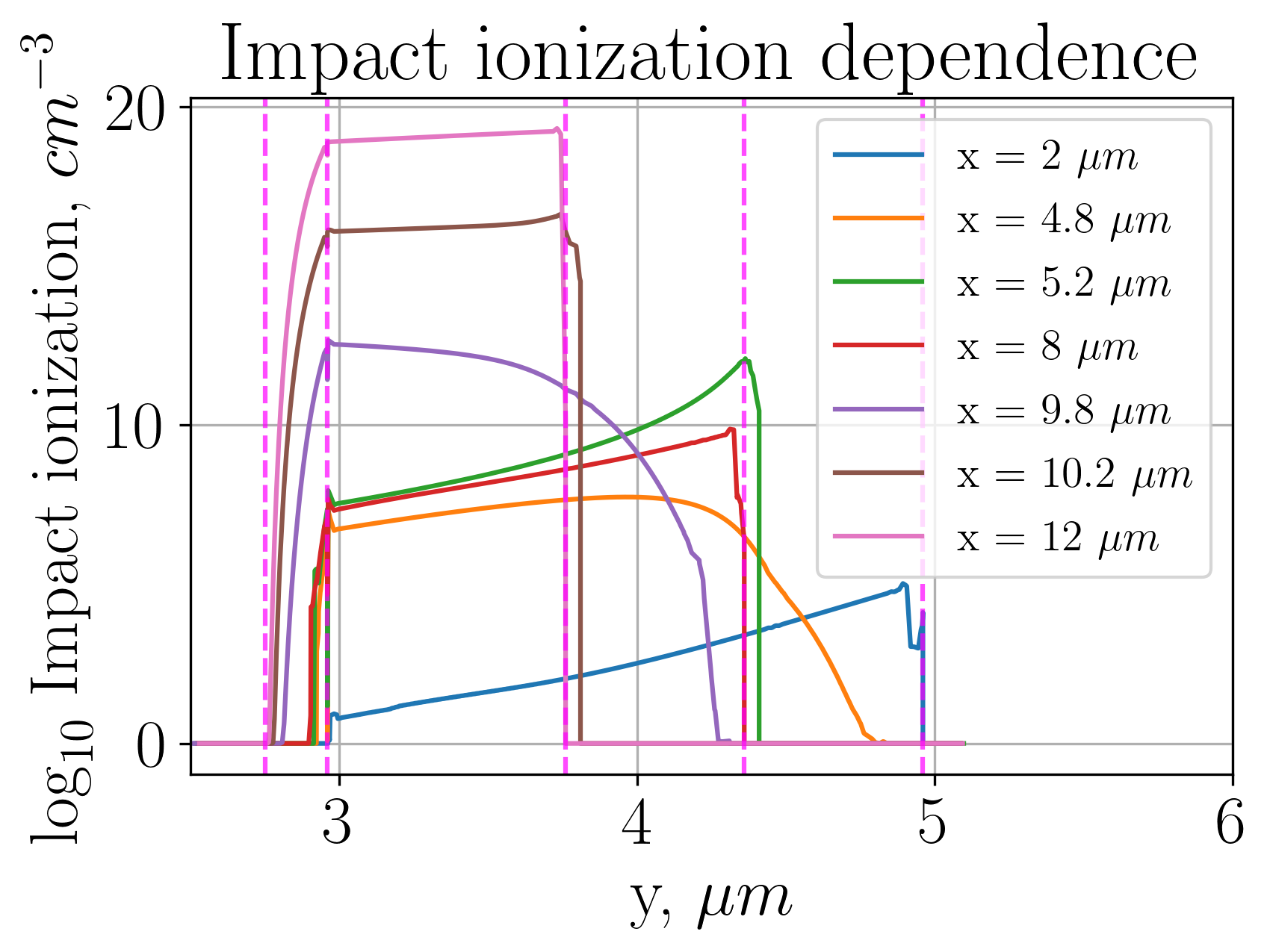}} d) \\
    \end{minipage}
    \caption{Three-level multiplication region structure with smooth transition with incident radiation intensity $1000 \ \mu W / cm^2$: a) heat map of electric field strength distribution; b) thermal map of avalanche generation rate distribution; c) electric field distribution profile in the cross-sections indicated in figure a); d) avalanche generation rate distribution profile in the cross-sections indicated in figure b).}
    \label{fig:3ur_smooth-3}
\end{figure}

Figures \ref{fig:3ur_smooth-5}, \ref{fig:3ur_smooth-4}, \ref{fig:3ur_smooth-3} demonstrate electric field distribution and profiles and avalanche generation rates for incident radiation with wavelength $\lambda = 1550$ nm and intensities $10 \ \ \mu W / cm^2$, $100 \ \ \mu W / cm^2$, $1000 \ \ \mu W / cm^2$, respectively.

These graphs were made for lower bias voltage than the plots without incident radiation.Therefore, the electric field strength in the multiplication region is much lower than in the previous graphs.

To investigate the avalanche generation rate more fully, the structure was fully irradiated, not just in the active region. In real devices, the radiation is concentrated in the active region, while less than $10 \%$ of the radiation reaches the inactive region. However, the proposed assumption allows us to better evaluate the feasibility of the presented structure.

In all the graphs shown, with structures illuminated in the active region, the avalanche generation rate is significantly higher than the avalanche generation rate in the region of local field enhancement. We believe that this is due to the recombination of carriers in this local region by means of SRH recombination processes. The avalanche generation rate in the region of intermediate thickness of the multiplication region is lower by about 8 orders of magnitude compared to the avalanche generation rate in the active region for all considered incident radiation intensities.

Such peculiarity of the considered structure allows applying it as a low noise avalanche photodiode (APD), i.e., operating it in linear mode.

The obtained distributions with incident radiation allow us to assess the peculiarities of the structure operation only in the linear mode. When operating in single-photon mode, only the curves without incident radiation must be considered.

\subsection{Simulation of devices with different active area diameters}

In this stage of the simulation, structures with three levels of multiplication region with a smooth transition were compared. The structures described above had an active region diameter of $D^{(25)}_{act} = 25 \ \ \mu m$. Next, we consider the structures with diameters $D^{(15)}_{act} = 15$ and $D^{(10)}_{act} = 10 \ \ \mu m$.

\begin{figure}[ht!]
    \centering
    \includegraphics[width=10cm]{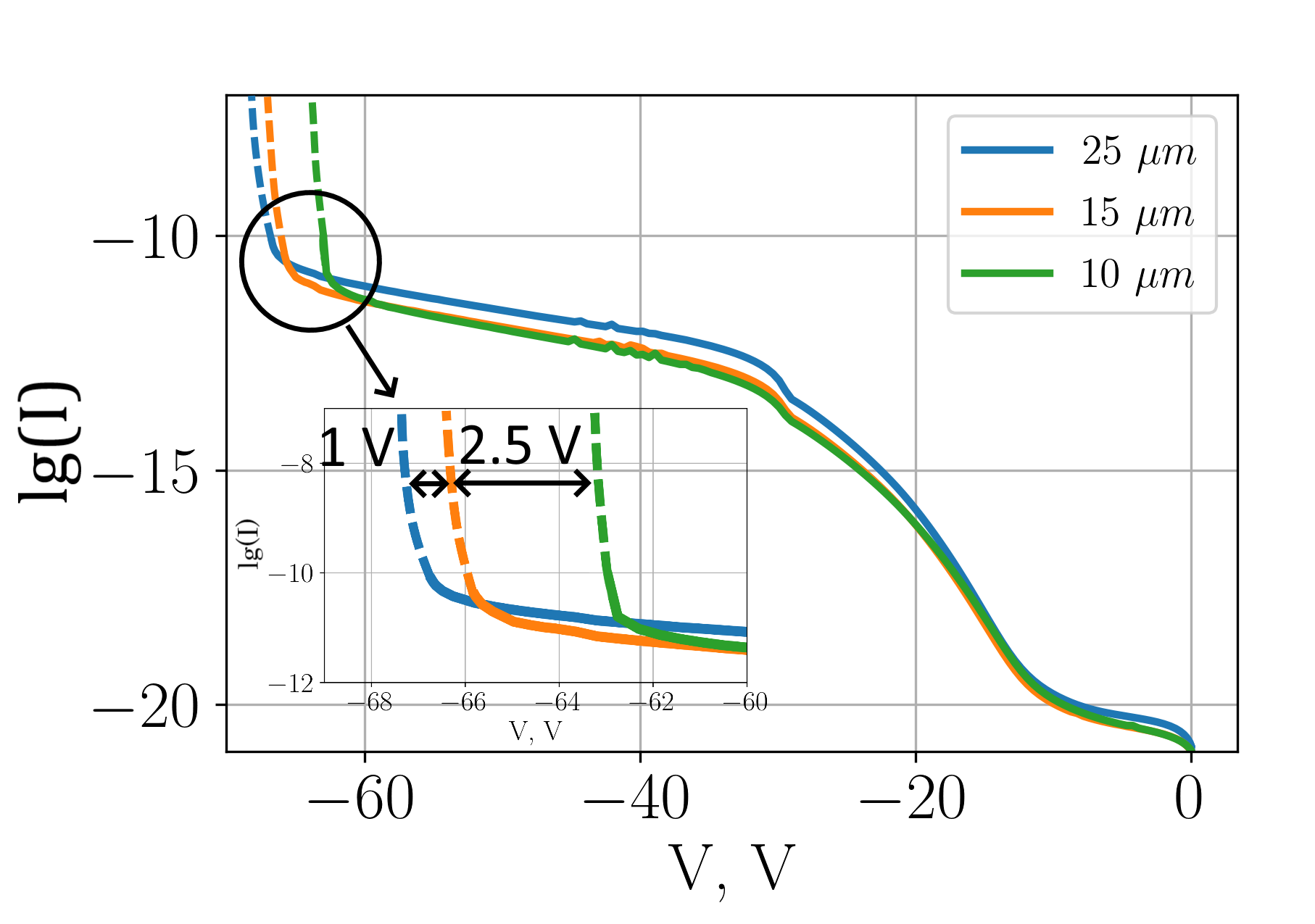}
    \caption{Comparison of the dark CVC for SPADs with a three-level structure of the multiplication region with smooth transitions for different diameters of the active region: $D = \{25, \ 15, \ 10 \} \ \mu m$.}
    \label{fig:ivcurve_square}
\end{figure}

The main advantage of using structures with a reduced active region diameter is a lower value of the dark current. On the other hand, the main drawback is the need for additional focusing of the radiation hitting the diode. In the simulation results presented, the CVCs of the dark current for diodes with different active region diameters have been compared (see Figure \ref{fig:ivcurve_square}).

As we can see in this figure, devices with a smaller active region diameter have a lower dark current in the linear part of the CVC. However, these devices also have a lower breakdown voltage. 

The dark current in the linear region of the $25 \ \ \mu m$ structure is higher than that of the $10 \ \ \mu m$ structure by approximately a factor of 3. The dark current in the linear region is nearly the same for the $15 \ \ \mu m$ and $10 \ \ \mu m$ diameter structures.

The breakdown voltage for the $25 \ \ \mu m$ diameter active region is  $1$ V higher than the value for the  $15 \ \ \mu m$ diameter structure, which in its turn is $2.5 \ \ V$ higher than the breakdown voltage of the $10 \ \ \mu m$ diameter structure. 

Let us consider the mechanism of dark current reduction in the linear region of the CVC.

The value of the active area for the discussed structures can be calculated by the circle area formula.

Thus, the expected reduction in the level of dark current in the active region should be proportional to the area of the active region. However, a greater contribution to the dark current is made by the dark current at transitions between levels of the multiplication region. The contribution of this quantity is proportional to the diameter of the active region. Assuming that the half-width of the transition region is $\Delta = 0.5 \ \ \mu m$, we can calculate the transition area as follows $S_{tr} = \dfrac{\pi}{4} ((D + \Delta)^2 - (D - \Delta)^2) = \pi D \Delta$.

We can also calculate the relative area as $s_{rel} = (S_{tr})/(S_{act}) = (\pi D \Delta)/( \dfrac{\pi}{4} D^2) = 4 \ \Delta/{D}$.

\begin{itemize}
    \item $25$ $\mu m$: $S_{act}^{(25)} \approx 490.9$ $\mu m$${}^2$ $\rightarrow$ $S^{(25)}_{tr} \approx 39.3$ $\mu m$${}^2$ $\rightarrow$ $s^{25}_{rel} \approx  0.08 = 8 \ \%$
    \item $15$ $\mu m$: $S_{act}^{(15)} \approx 176.7$ $\mu m$${}^2$ $\rightarrow$ $S^{(15)}_{tr} \approx 23.6$ $\mu m$${}^2$ $\rightarrow$ $s^{15}_{rel} \approx  0.013 = 13 \ \%$
    \item $10$ $\mu m$: $S_{act}^{(10)} \approx 78.5$ $\mu m$${}^2$ $\rightarrow$ $S^{(10)}_{tr} \approx 15.7$ $\mu m$${}^2$ $\rightarrow$ $s^{10}_{rel} \approx  0.2 = 20 \ \%$
\end{itemize}

The general equation for the dark current $I_{dcr}$ for this structure can be written as:

\begin{equation}
    I_{dcr} = S_{act}  j_{act} + S_{tr}  j_{tr},
\end{equation}

\noindent where $j_{act}$ -  current density in the active region; $j_{tr}$ -  current density in the transition region.

Given that $S_{act}$ is related to $S_{tr}$, this equation can be rewritten as:

\begin{equation}
    I_{dcr} = j_{act} \dfrac{\pi}{4} (D^2 + 4 k D \Delta) = j_{act} \dfrac{\pi}{4} D^2 (1 + 4 k \dfrac{\Delta}{D}).
\end{equation}

\begin{figure}[ht!]
    \begin{minipage}[h]{0.47\linewidth}
    \center{\includegraphics[width=1\linewidth]{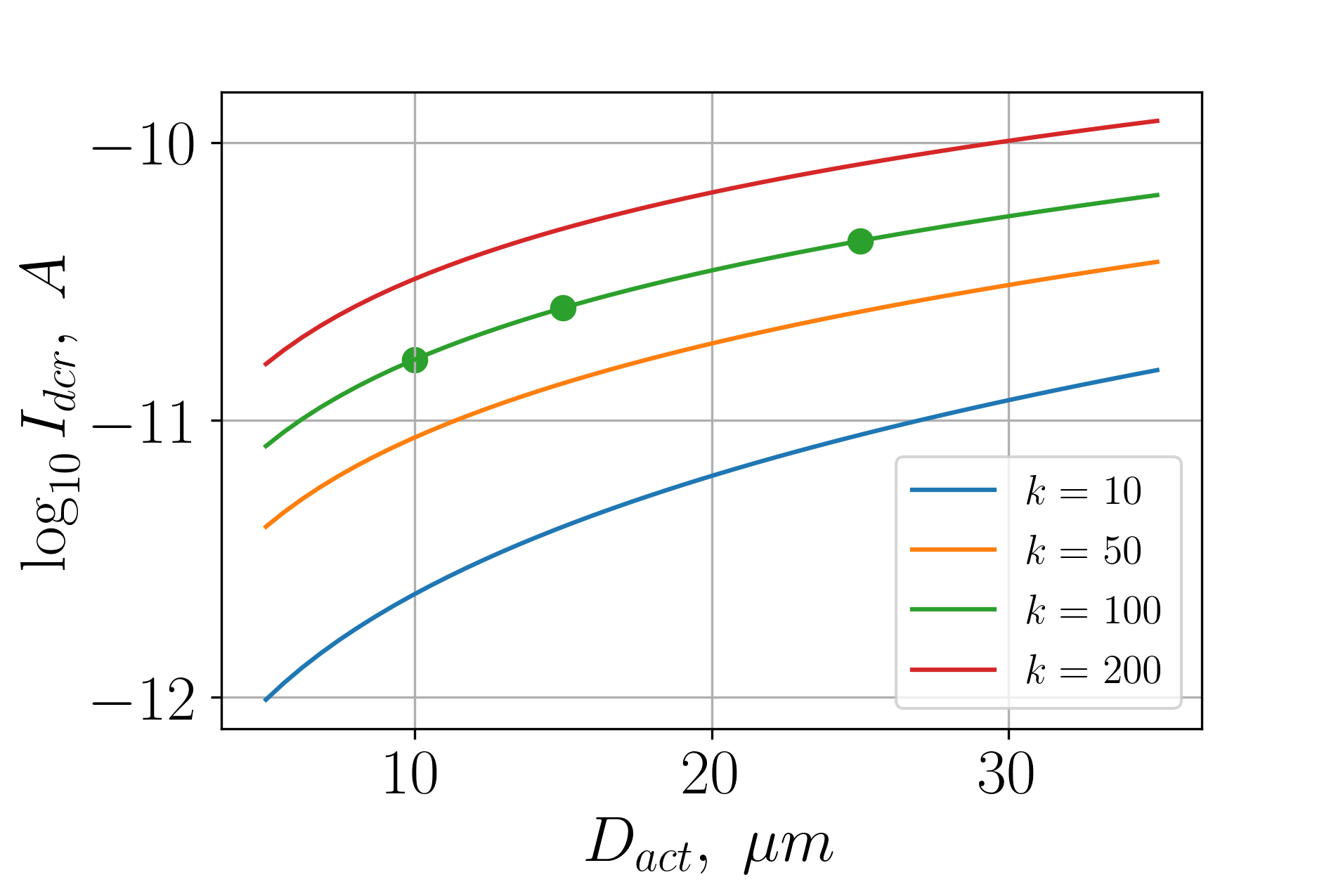}} a) \\
    \end{minipage}
    \hfill
    \begin{minipage}[h]{0.47\linewidth}
    \center{\includegraphics[width=1\linewidth]{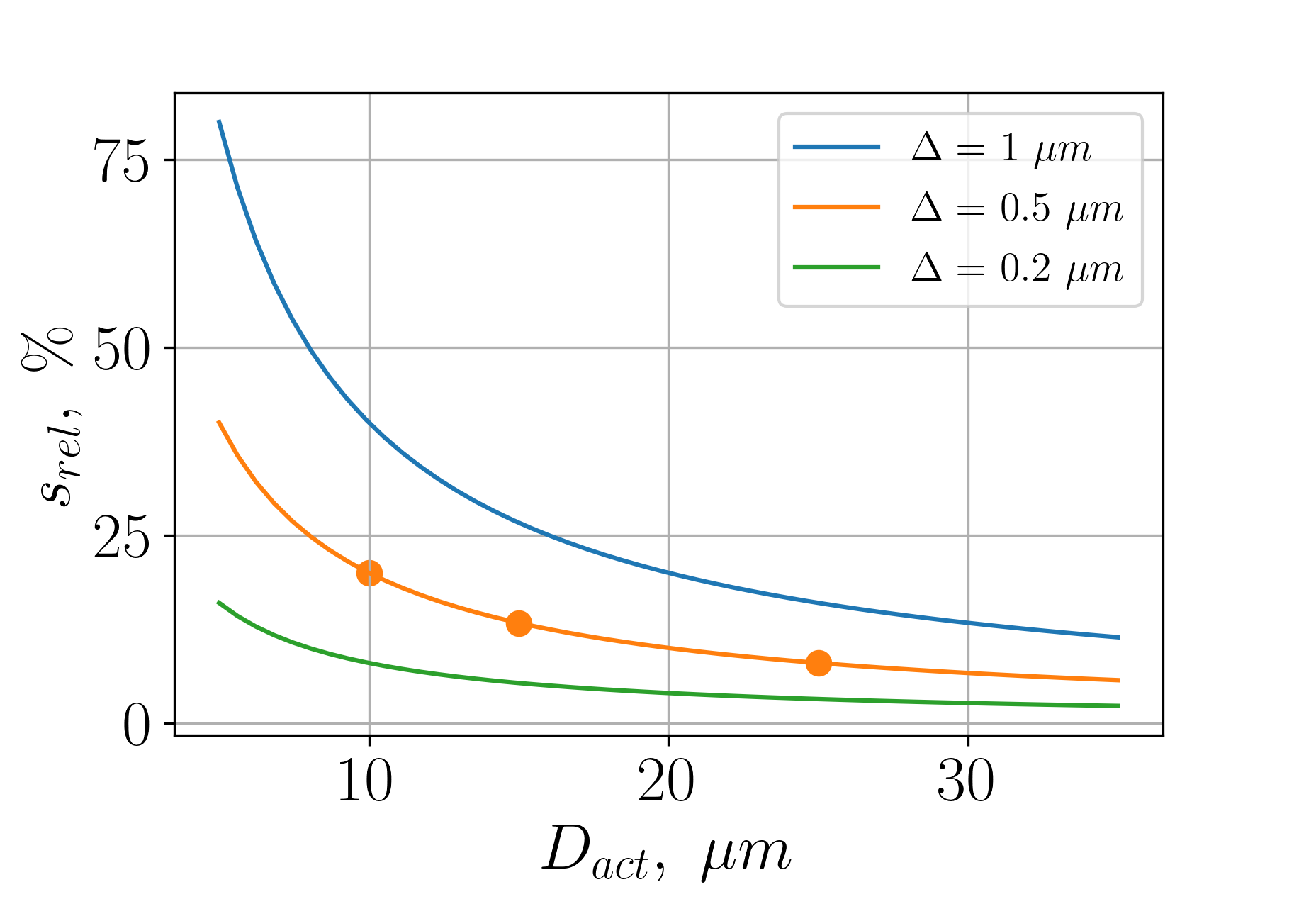}} \\b)
    \end{minipage}
    \caption{a) Dependence of the dark current on the diameter of the active region for different ratios of the current density in the transition region and the active region. b) Dependence of the relative area $s_{rel}$ on the diameter of the active region at different values on the width of the transition region in the multiplication region.}
    \label{fig:dcr_diam_analyt}
\end{figure}

Thus, the greater the ratio $\dfrac{\Delta}{D}$, i.e. the smaller the diameter $D$ for a given $\Delta$, the greater the influence of the transition region on the dark current. Let us assume that the current density $j_{act}$ in the active region does not depend on the area of the active region. The dependence of the dark current on the diode diameter, in this case, is shown in Figure \ref{fig:dcr_diam_analyt} a). The dots mark the device parameters used in the simulation.

We can see from these figures that by reducing the diameter of the active region from $25 \ \ \mu m$ to $10 \ \ \mu m$ we can reduce the dark current level by approximately $8.5$ dB. In Figure \ref{fig:ivcurve_square} we consider the linear region of the CVC. We can see that the reduction in dark current is in good agreement with theoretical predictions.

The nature of the breakdown voltage reduction by a decrease of the diode diameter is related to changes in the relative sizes of the active and transition areas in the multiplication region (specifically, the increase of relative area $s_{rel}$). An increase in $s_{rel}$ results in a larger relative fraction of the structure having a smaller electric field boundary for avalanche breakdown. Thus, the larger the relative area, the lower the breakdown voltage for the same value of the transition region width.

Figure \ref{fig:dcr_diam_analyt} b) shows the dependence of $s_{rel}$ on the diameter of the active region for different values of the transition region width in the multiplication region. The points correspond to the parameters at which the simulation was performed.

A three-level multiplication region with smooth transitions provides a reduction of electric field inhomogeneities and, consequently, avalanche generation at the boundaries between the levels of the multiplication region. It should be noted that the formation of smooth transitions in the multiplication region is more technological than growing a multilevel structure using other approaches. In this design, the electric field is localized predominantly in the active region of the multiplication region, resulting in a lower dark current in the linear section of the CVC and $DCR$ when operating the diode in the Geiger mode. This also means that decreasing the diameter of the active region of the multiplication region allows reducing the zone of the electric field localization responsible for avalanche generation, which leads to an even greater decrease of $DCR$ and dark current.

\section{Results and discussion}

In this section, we compare the presented structure types based on the analysis of the CVC.

\begin{figure}[h]
    \centering
    \includegraphics[width=10cm]{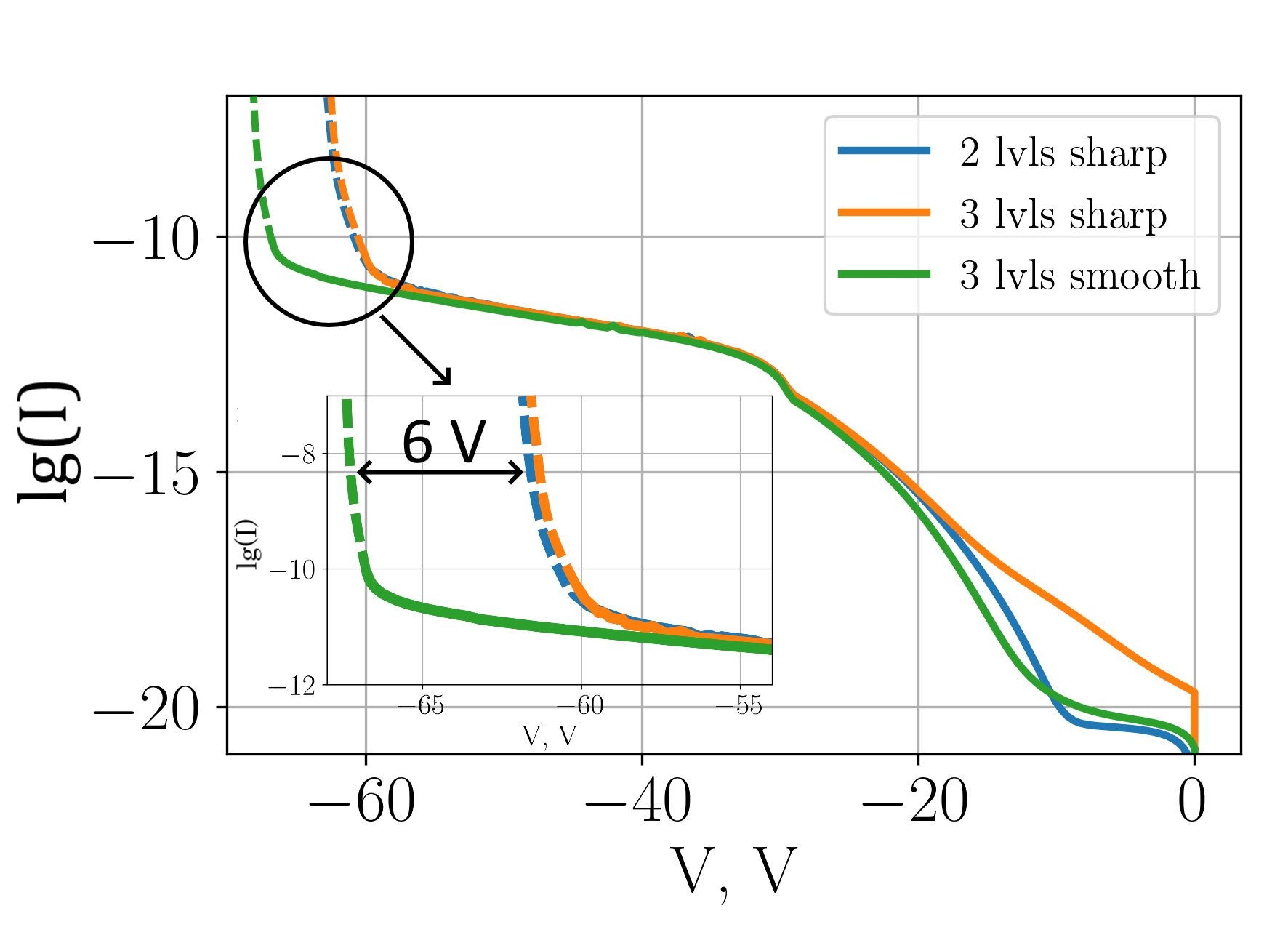}
    \caption{Comparison of the dark current for SPADs with different types of structures.}
    \label{fig:ivcurves_comparedcr}
\end{figure}

Figure \ref{fig:ivcurves_comparedcr} shows the CVCs for different types of structures without incident radiation. The avalanche breakdown for a structure with a three-level multiplication region with sharp transitions occurs about $200$ mV earlier than for a structure with a two-level multiplication region with sharp transitions. The avalanche breakdown voltage fully determines the limits of device operation in Geiger and linear modes.

In the considered structures, avalanche breakdown occurs primarily in transitions between different levels of the multiplication region. Therefore, the higher the local voltage at the transition, the lower the breakdown voltage value of the structure. The plots of CVC and breakdown voltage quantitatively express the result of the combined effects of generation and recombination and allow us to assess the quality of the created structure. The primary electron (photon- or noise-generated) in the active region of the structure will have an extremely low probability of generating an avalanche process because the local breakdown voltage in the active region of the structure will be much higher than that applied to the SPAD. However, by all operation indicators, it will appear to work in the Geiger mode due to avalanche breakdown in transitions between the multiplication region levels. Such a device would therefore have an extremely low $PDE$. We can conclude that the smaller the local increase of field strength in the transitions, the higher $PDE$ the SPAD would have in operation. This effect can be diagnosed by the CVC of the device.

\begin{figure}[h]
    \centering
    \includegraphics[width=10cm]{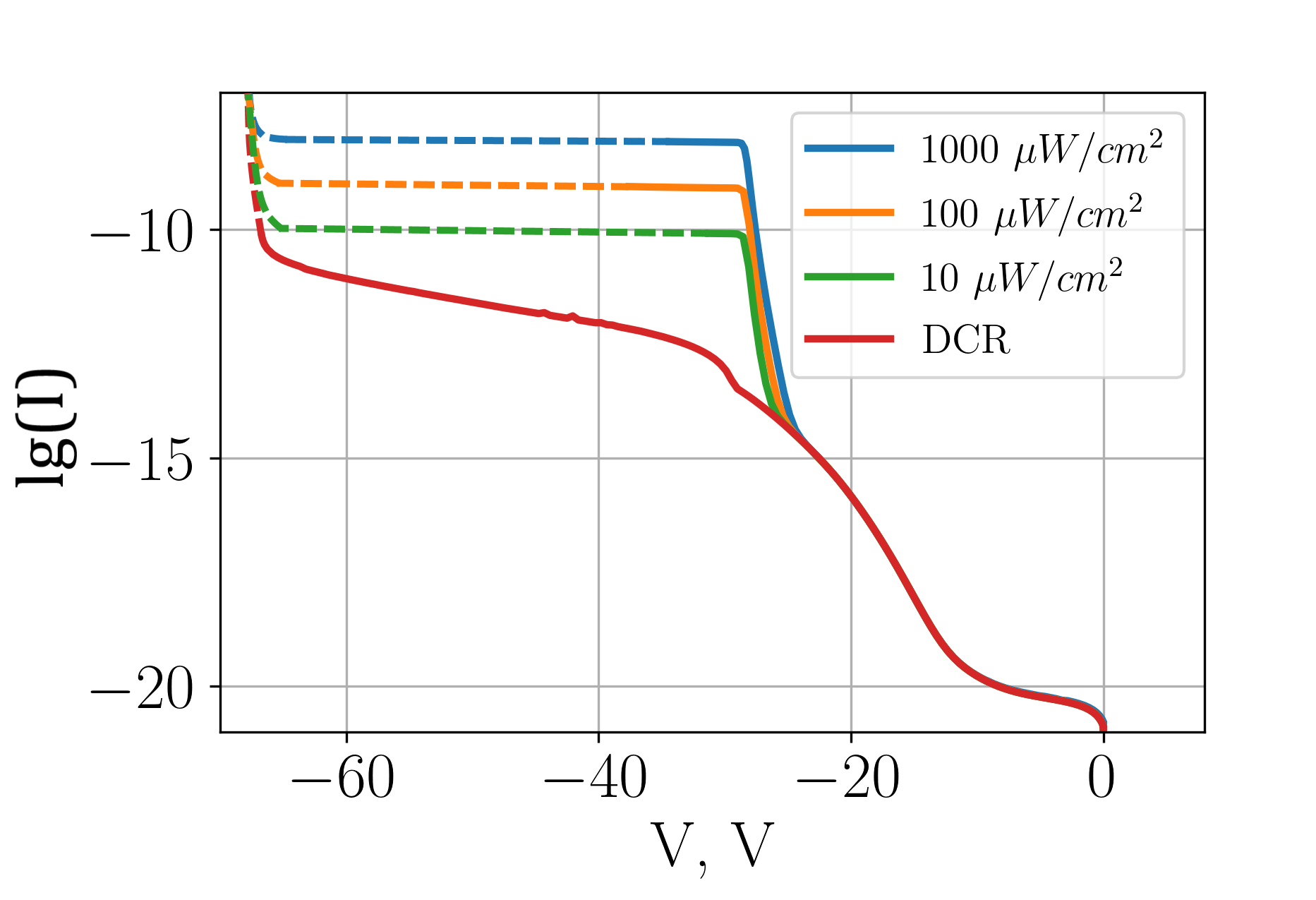}
    \caption{Comparison of CVC for different incident radiation intensities for SPADs with a three-level multiplication structure with smooth transitions.}
    \label{fig:ivcurves_v3struct_red}
\end{figure}

The $6$ V difference between the breakdown voltage values of the considered structures clearly shows that the $PDE$ of the SPAD based on the structure with three levels of the multiplication region and a smooth transition (see Figure \ref{fig:2d_struct_3smooth}) will be significantly higher than that of the SPAD based on the other presented structures. This once again proves that the creation of smooth transitions between different levels of the multiplication region is an important aspect of SPAD structure optimization.

Figure \ref{fig:ivcurves_v3struct_red} shows CVC plots for different incident radiation intensities for a SPAD with a three-level multiplication structure with a smooth transition. This graph shows that an order of magnitude change in incident radiation intensity results in order of magnitude change in current, at least in the range of $10 ... 1000 \ \mu W /cm^2$. Thus, the device based on the structure presented in this paper can be effectively used not only in the Geiger mode but also in the linear mode, such as the APD.

\section{Conclusion}

The simulation and analysis of electric field profiles and avalanche generation rates in different structures have shown that the creation of smooth transitions at the level boundaries of the multiplication region is an extremely effective method of $DCR$reduction in SPAD. If one manages to achieve sufficiently smooth transitions (see Figure \ref{fig:2d_struct_3smooth}) in the device design so that the dark count rate increase due to local field enhancement does not exceed $20$ dB above the active region value, a three-level multiplication region can be considered for implementation.

It has been shown that changing the diameter of the active region of the device results in both a change in the dark current in the linear portion of the CVC and a change in the breakdown voltage. It has been shown that decreasing the diameter from $25 \ \ \mu m$ to $10 \ \ \mu m$ reduces the dark current in the linear operation mode of the device by approximately $10$ dB. However, it results in the breakdown voltage decrease by approximately $3.5$ V. The first property leads to a decrease in $DCR$ when the device is operated in the Geiger mode. However, the second property reduces the probability of photon detection.

It has been shown that the quality of the SPAD device can be assessed by the avalanche breakdown voltage and the overall CVC plot, provided the examined structures have equal thicknesses of the multiplication region and other layers of the structure. The higher the breakdown voltage, the better the structure's quality due to smaller local increases in the field strength. Following this statement, we conclude that for further use in single-photon detectors, it is reasonable to pick specific SPADs from a batch on the sole basis of their current-voltage curves.

The developed SPAD structure with three levels of multiplication region and smooth transitions can also be used for APD devices operating in linear mode. Such a device has the following property: when the intensity increases by one order of magnitude, the current changes by one order of magnitude as well, which is beneficial for the linearity of the characteristic. In addition, this device is characterized by low dark current and, therefore low signal-to-noise ratio.

\begin{backmatter}

\bmsection{Funding}
The Program of Strategic Academic Leadership “Priority 2030” (Contract № L-2022-SP2-P02-027 from 01.07.2022)

\bmsection{Acknowledgments}
The Ministry of Education and Science of the Russian Federation in the framework of the Program of Strategic Academic Leadership “Priority 2030” (Strategic Project “Quantum Internet”, Grant № K1-2022-027)

\bmsection{Disclosures}
The authors declare no conflicts of interest.

\bmsection{Data Availability Statement} 
No data were generated or analyzed in the presented research.

\end{backmatter}

\bibliography{bibliography}

\end{document}